\documentclass{article}
\usepackage{arxiv}
\usepackage[utf8]{inputenc} 
\usepackage[T1]{fontenc}    
\usepackage[dvipsnames]{xcolor}
\usepackage{hyperref}       
\newcommand\myshade{85}
\colorlet{mylinkcolor}{violet}
\colorlet{mycitecolor}{YellowOrange}
\colorlet{myurlcolor}{Aquamarine}

\hypersetup{
  linkcolor  = mylinkcolor!\myshade!black,
  citecolor  = mycitecolor!\myshade!black,
  urlcolor   = myurlcolor!\myshade!black,
  colorlinks = true,
}

\usepackage{booktabs}       
\usepackage{amsfonts}       
\usepackage{nicefrac}       
\usepackage{microtype}      
\usepackage{lipsum}
\usepackage{graphicx}
\usepackage{amssymb}
\usepackage{bm}
\usepackage{amsmath}
\usepackage[biblabel]{cite}
\usepackage[bottom]{footmisc}

\title{FusionAccel: A General Re-configurable Deep Learning Inference Accelerator on FPGA for Convolutional Neural Networks}

\author{
  Shi ~Shi\\
  Fudan University\\
  Shanghai, 201203 \\
  \texttt{sshi15@fudan.edu.cn} \\
}

\begin{document}
\maketitle

\begin{abstract}
The deep learning accelerator is one of the methods to accelerate deep learning network computations, which is mainly based on convolutional neural network acceleration. To address the fact that concurrent convolutional neural network accelerators are not solely open-source and the exclusiveness of platforms, FusionAccel, a scalable convolutional neural network accelerator hardware architecture with supporting software is proposed. It can adapt to different network structures and can be reconstructed before compilation and reconfigured at runtime. This paper realizes this RTL convolutional neural network accelerator design and functional verifications on a Xilinx Spartan-6 FPGA. The result is identical to that of Caffe-CPU. Since the entire project is based on RTL, it can be migrated to ASIC after replacing some FPGA-specific IPs.
\end{abstract}


\section{Introduction}
The applications of deep learning, such as image recognition, autonomous driving, and smart robots, are significantly changing people's lives. After CNNs (convolutional neural networks) are proposed, the accuracy of image recognition and the capability of video classification improve significantly. Meanwhile, the networks are getting bigger and deeper to extract more raw information from images and videos. From AlexNet in 2012 to ResNet in 2015, the model size has increased by 16 times \cite{han2017efficient}.

To improve the efficiency of deep learning computations, methods like pruning, weight sharing, quantization and accelerators are proposed in recent years\cite{han2017efficient}. Pruning eliminates some connections of nodes in a network, while maintains the forwarding accuracy of the network at a similar level, reducing the computation overhead during training and inference. Weight sharing clusters the neighboring weights and generates a corresponding lookup table, reducing the storage overhead of network models. Quantization maps network weights of higher precision to lower precision and retrains the whole network, while the accuracy is not compromised. Accelerator computes convolutions or multi-channel operations in parallel or in batch with parallel algorithms and specific hardware, reducing memory access times and elapse time in inference.

GPUs (Graphics Processing Unit) are one type of general deep learning accelerators, while deep learning accelerators discussed in this paper refer to those deployed on ASICs (Application Specific Integrated Circuit) or FPGAs (Field Programmable Gate Arrays).

\section{Concurrent open-source projects}
With regards to accelerator technology, there are two research concentrations: high-performance \cite{iandola2016squeezenet, gokhale2014nn, gholami2018squeezenext,wu2017squeezedet, han2016eie, cavigelli2016origami} and re-configurability \cite{kestur2012towards, farabet2009cnp, farabet2011neuflow, gysel2016hardware}. ASIC accelerators tend to make use of the performance advantage and higher data bandwidth, while FPGA ones tend to make use of the configurability to support more network types. Concurrent two large open-source FPGA accelerators are NVDLA and CHaiDNN.

\subsection{NVDLA by NVIDIA}
\begin{figure}
  \centering
  \includegraphics[width=0.8\textwidth]{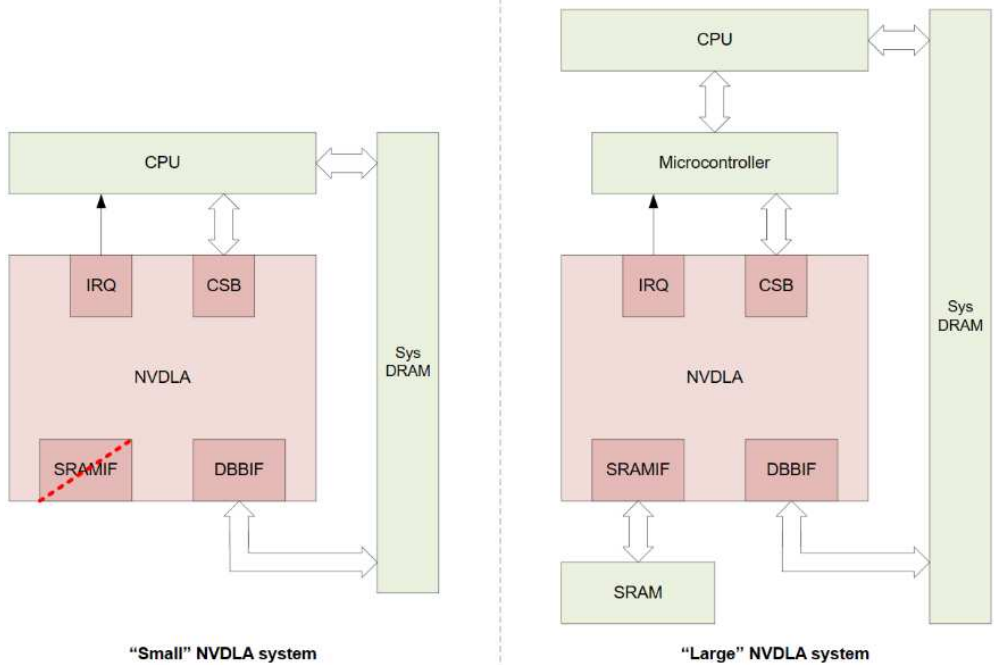} 
  \caption{NVDLA system block diagram.}
  \label{fig:fig1}
\end{figure}

NVDLA \cite{nvdla} is an open-source deep learning accelerator released by NVIDIA. Its source code is based on HDL (hardware description language). It can be either deployed on FPGAs, SoCs (system on chip, e.g., Xilinx ZYNQ series) or ASICs. When it is deployed on FPGAs, the accelerator communicates with the host memory with high-speed buses like PCIe and CAPI, controlled by CPUs like X86 and POWER. When it is deployed on SoC, it can access memories with on-chip AXI interfaces, controlled by CPUs like ARM. NVDLA's main design feature is the modularized, re-configurable and scalable architecture. It supports inference precisions of INT8, FP16 and FP32. The architecture is as Figure \ref{fig:fig1} \cite{nvdla}. The red part is the NVDLA hardware. CSB (Configuration Space Bus) gets commands from CPU. DBBIF (Data Backbone Interface) transfers the memory data shared with system DRAM. IRQ (Interrupt Interface) sends interrupt signals to CPU. In NVDLA large systems, SRAMs cache data to speed up data access.

\begin{figure}
  \centering
  \includegraphics[width=0.6\textwidth]{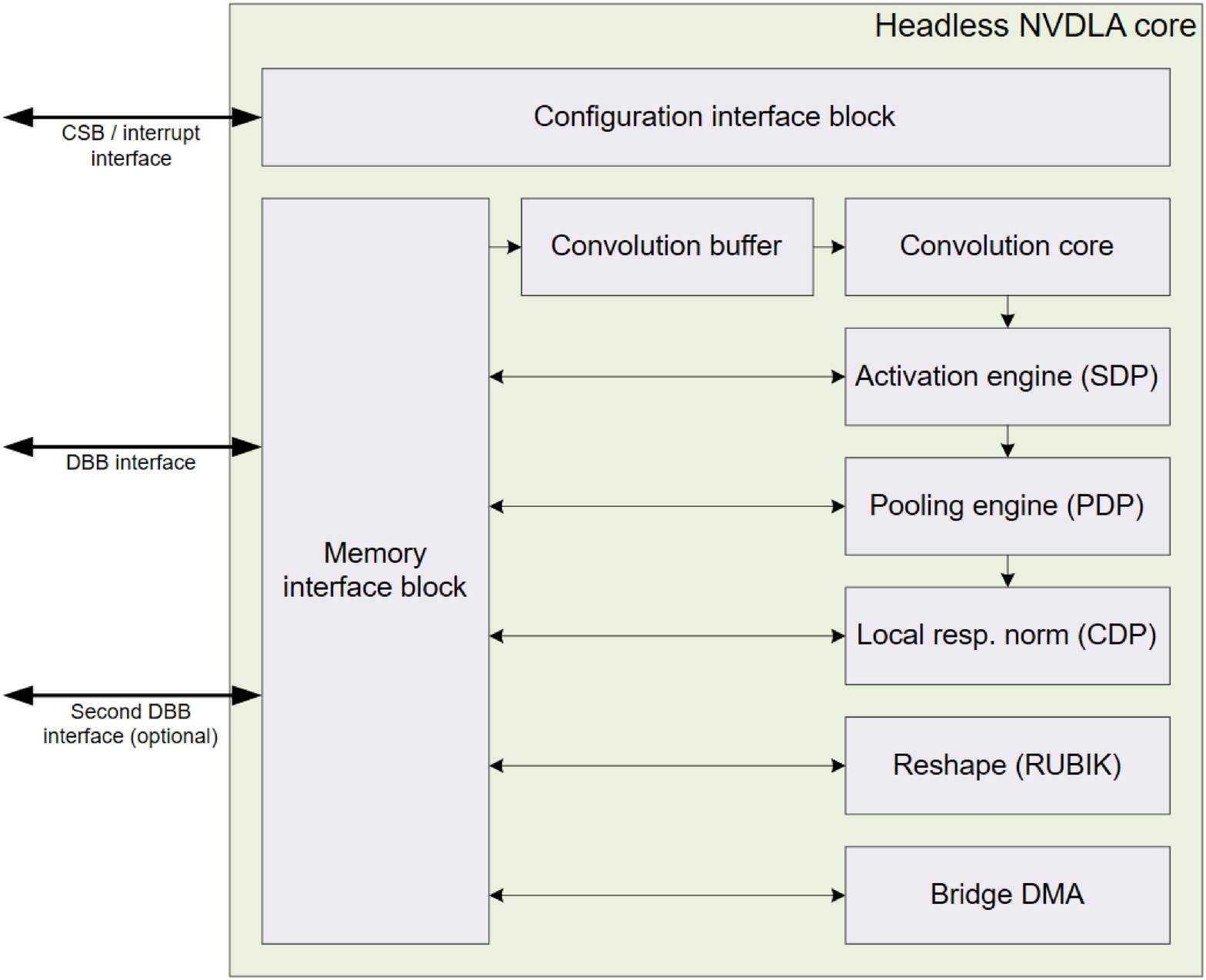} 
  \caption{NVDLA inner core.}
  \label{fig:fig2}
\end{figure}

Figure \ref{fig:fig2} \cite{nvdla} depicts the inner core of NVDLA. The computation unit consists of the following parts: Convolution buffer and convolution core, Activation engine (SDP, Surface Data Processor), Pooling engine (PDP, Planar Data Processor), LRN engine (CDP, Channel Data Processor), Memory Reshaping Unit (RUBIK) and bridge DMA (Direct Memory Access). The commands required by the core are transferred via the CSB interface. The data is transferred via the bridge DMA interface.

However, the open-source NVDLA project supports only pre-generated loadable files and restricted network types in the software section, while its parser, compiler and optimizer \cite{nvdla} are not open-source. Users can not put other networks or self-trained networks on it.

\begin{figure}
  \centering
  \includegraphics[width=1.0\textwidth]{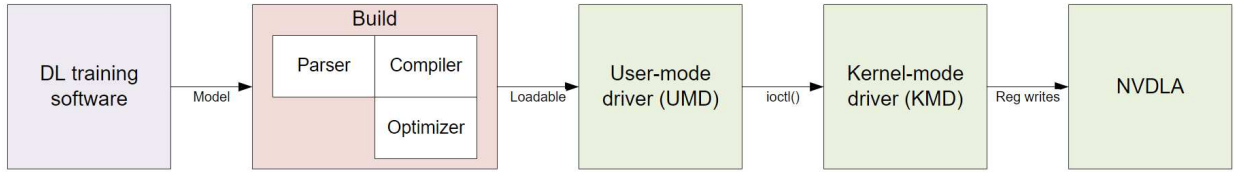} 
  \caption{NVDLA software flow.}
  \label{fig:fig3}
\end{figure}

\subsection{CHaiDNN by Xilinx}
CHaiDNN \cite{chaidnn} is an open-source deep learning inference accelerator released by Xilinx. The source code is written in C++, so it can only be deployed on Xilinx ZYNQ MPSoCs or UltraScale MPSoCs. The main feature of CHaiDNN is that it achieves a balance between computation efficiency and accuracy with 6-bit or 8-bit quantized data. On Xilinx ZU9/ZU7 platforms, a maximum of 1024/512 on-chip DSPs can be utilized. Apart from NVDLA, the fully connected layer, which contains the most weights in CNNs, is realized on CPUs. CHaiDNN project is compiled with Xilinx SDSoC. The SDSoC toolkit analyzes and makes partitions of the C/C++ code. The parts that can be accelerated by the parallel resources on FPGAs will be generated as programmable logic, while the parts like memory management, interrupt control, sequential flow control will be generated for ARM CPUs in the SoC to maximize the efficiency. However, due to its platform exclusiveness, this project cannot be directly migrated onto ASICs.

\section{Convolutional Neural Networks, Parallel Optimizations and Corresponding Hardware Relationships}

\subsection{Hardware Platform}
\begin{figure}
  \centering
  \begin{minipage}[t]{0.48\textwidth}
  \centering
  \includegraphics[width=0.64\textwidth]{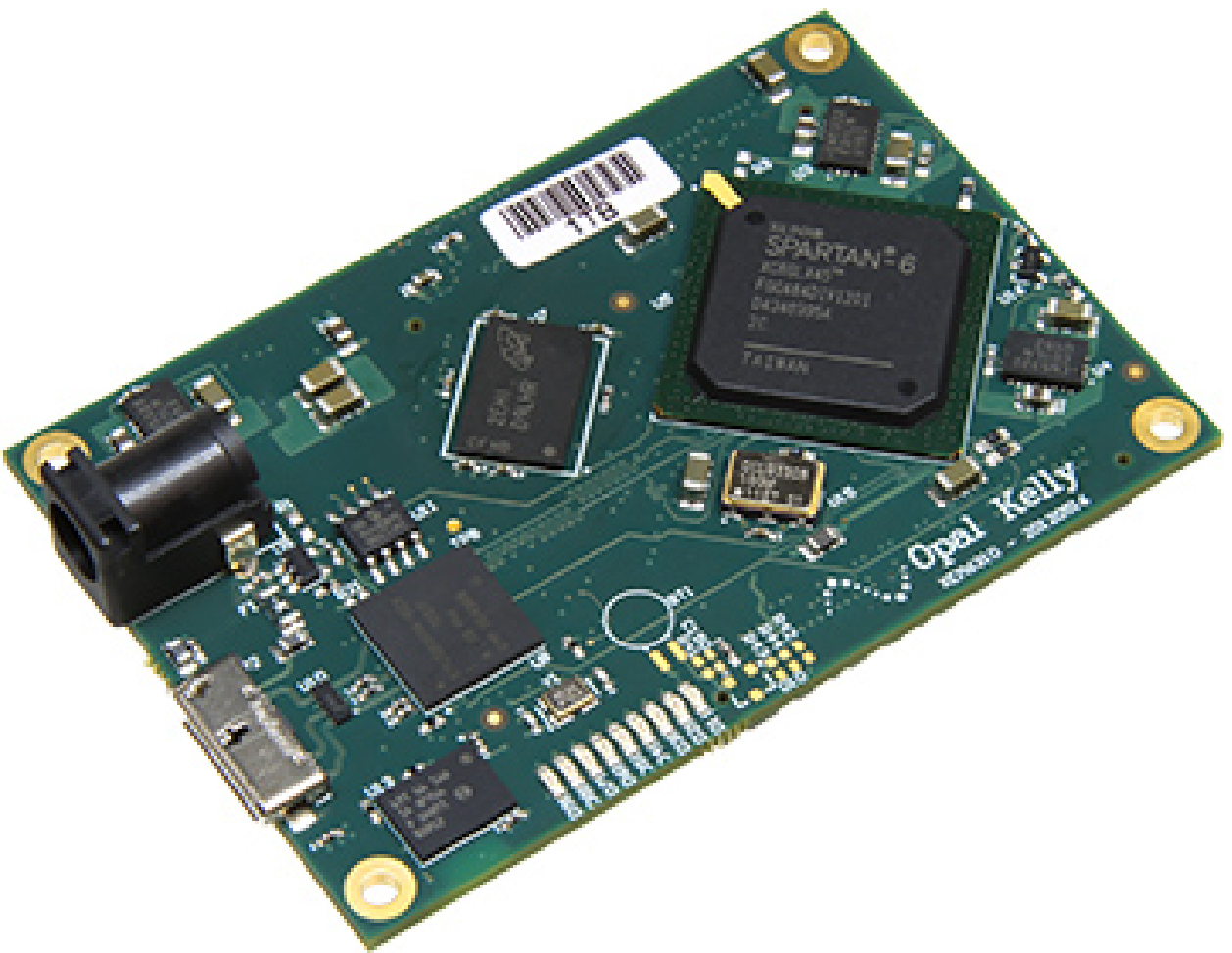} 
  \caption{Opal Kelly XEM6310-LX45 FPGA.}
  \label{fig:fig4}
  \end{minipage}
  \begin{minipage}[t]{0.48\textwidth}
  \centering
  \includegraphics[width=0.55\textwidth]{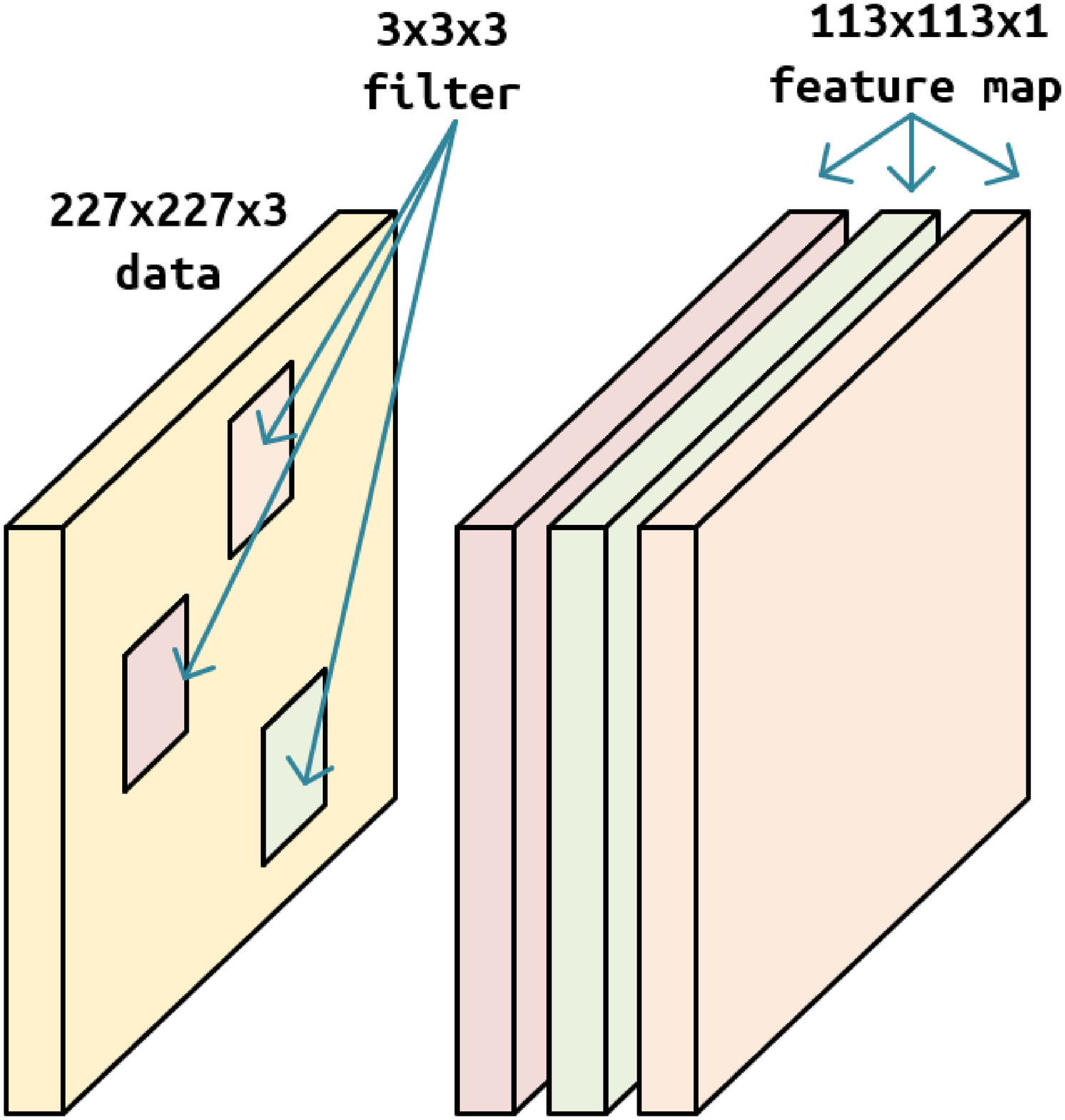} 
  \caption{Convolution diagrams.}
  \label{fig:fig5}
  \end{minipage}
\end{figure}

Due to the demand of ASIC migration, the hardware of this paper is FPGA. The model is Opal Kelly XEM6310-LX45 and the chip is 45nm Xilinx Spartan-6 XC6SLX45-2, as Figure \ref{fig:fig4} \cite{xem6310}. The available hardware resources are as follows: In terms of computation resource, there are 58 Mult/DSPs and 6822 slices (each slice contains 4 6-LUTs and 8 DFFs). In terms of storage, there are totally 2088Kbit BRAMs (Block Memory) and a 1Gbit DDR2 RAM (16 bits wide, 10Gb/s bandwidth)\footnote{\href{http://assets00.opalkelly.com/library/XEM6310-UM.pdf}{Opal Kelly XEM6310 User Manual}. Accessed March 6, 2019.}.

Moreover, another reason to develop on this FPGA is that it has high speed USB3.0 connection and corresponding USB3.0 chip on-board. It can achieve a maximum transfer speed of 340MB/s, which is fast to load block data. Meanwhile, Opal Kelly provides FrontPanel HDL APIs so that users can program with C/C++, C\#, Ruby, Python and Java\cite{xem6310}.

\subsection{Characteristics of Convolutional Neural Network Computation}
CNN inference normally consists of the following layers: convolution, pooling, activation and local response normalization. LRN layer is proposed in AlexNet, but it seldom appears in recent CNN researches, and networks without it can achieve a same accuracy of inference as those with it. Moreover, there are AlexNet and GoogLeNet without LRN layers proposed, so LRN layers are not realized in this work.

Figure \ref{fig:fig5} shows the convolution layer operation. The input matrix dimension is $w\times h\times c\ (w = h)$, the $k\times k\times c$ input data matrix and $k\times k\times c$ weight matrix in the sliding window are dot-multiplied and summed. The result is a corresponding value in the output matrix. The stride of each sliding operation is s. To achieve an exact division, padding size of p is required for the input matrix surface. Thus, the final output surface side is $w' = (w - k + 2p)/s + 1$. To extract different types of information, normally there are multiple (n) weight cores to slide on the same input data matrix. Thus, the final output matrix dimension is $w'\times h'\times n\ (w'=h')$. n increases as the network forwarding goes on, which means that the more information extracted, the deeper the channel dimension becomes. Different weight cubes have different bias values. In networks like AlexNet there are fully connected layers, which are essentially 1x1 convolutions, so fully connected layers are merged to convolutional layers. Fully connected layers require tremendous weights, which account for the main part of convolution operation. Obviously, convolution layers account for the main part of the network operation.

\begin{figure}
  \centering
  \begin{minipage}[c]{0.9\textwidth}
  \includegraphics[width=0.9\textwidth]{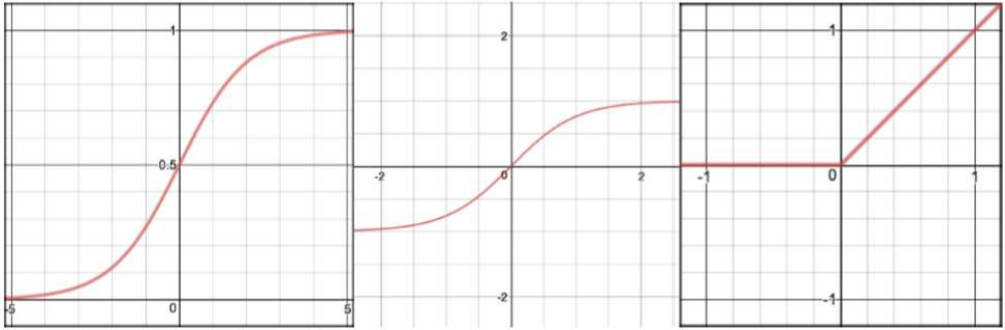} 
  \end{minipage}
  \caption{Activation Function (sigmoid, tanh, ReLu).}
  \label{fig:fig6}
\end{figure}

The activation Layer maps each element of the input matrix to another value. It does not change the matrix dimension but serves to improve the network comprehension through non-linear transformation. Figure \ref{fig:fig6} shows common activation functions like sigmoid, tanh, ReLu. ReLu is used most commonly, as its computation overhead is the smallest and it can help the network achieve a fast convergence during training. In terms of hardware implementation, ReLu is also the easiest since it is only required to judge the sign bit of a floating-point number.

\begin{figure}
  \centering
  \includegraphics[width=0.9\textwidth]{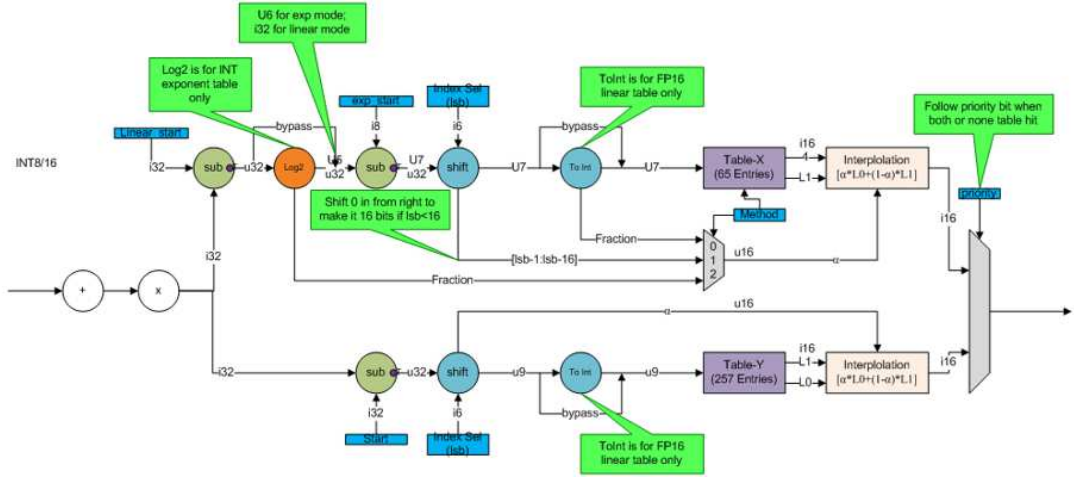} 
  \caption{NVDLA lookup table structure.}
  \label{fig:fig7}
\end{figure}

\begin{figure}
  \centering
  \begin{minipage}[c]{0.45\textwidth}
  \centering
  \includegraphics[width=0.9\textwidth]{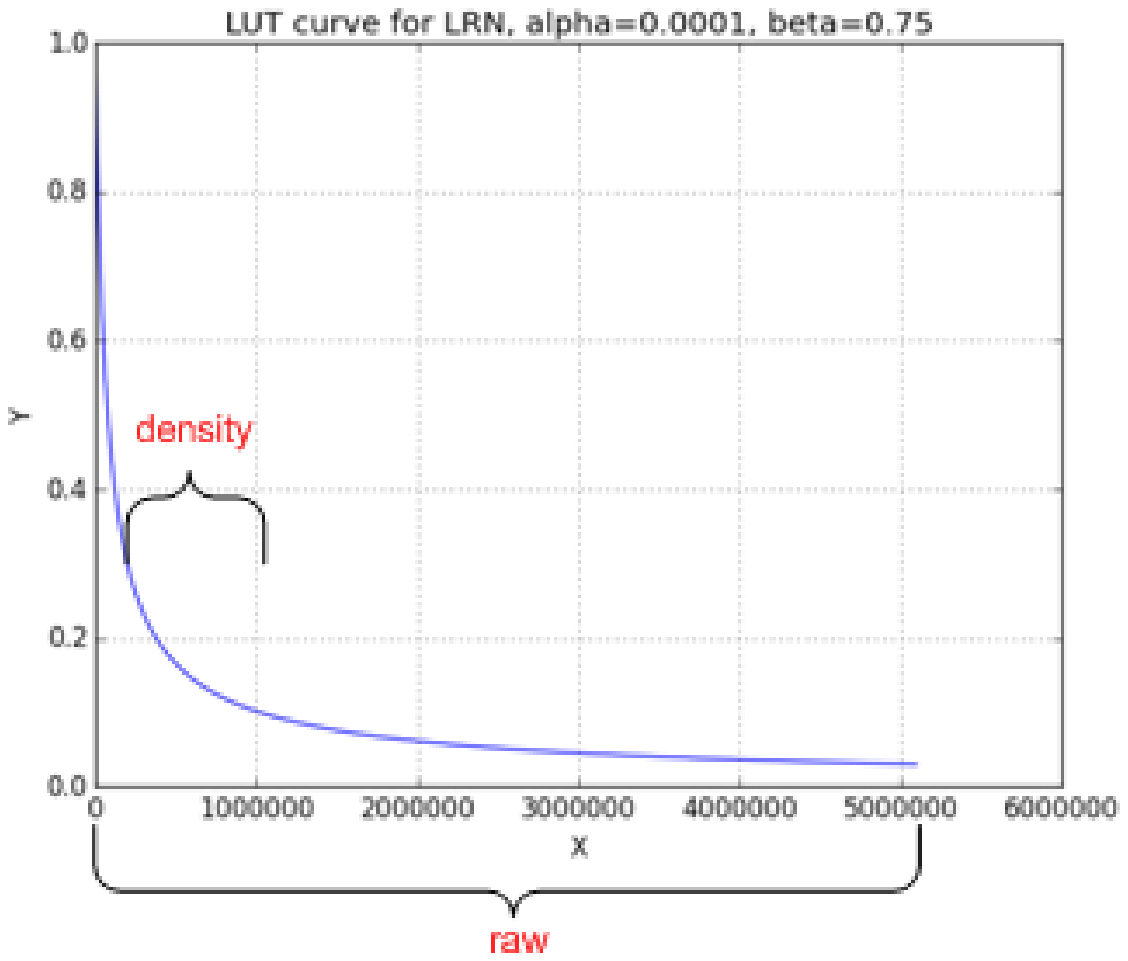} 
  \end{minipage}
  \begin{minipage}[c]{0.45\textwidth}
  \centering
  \includegraphics[width=0.9\textwidth]{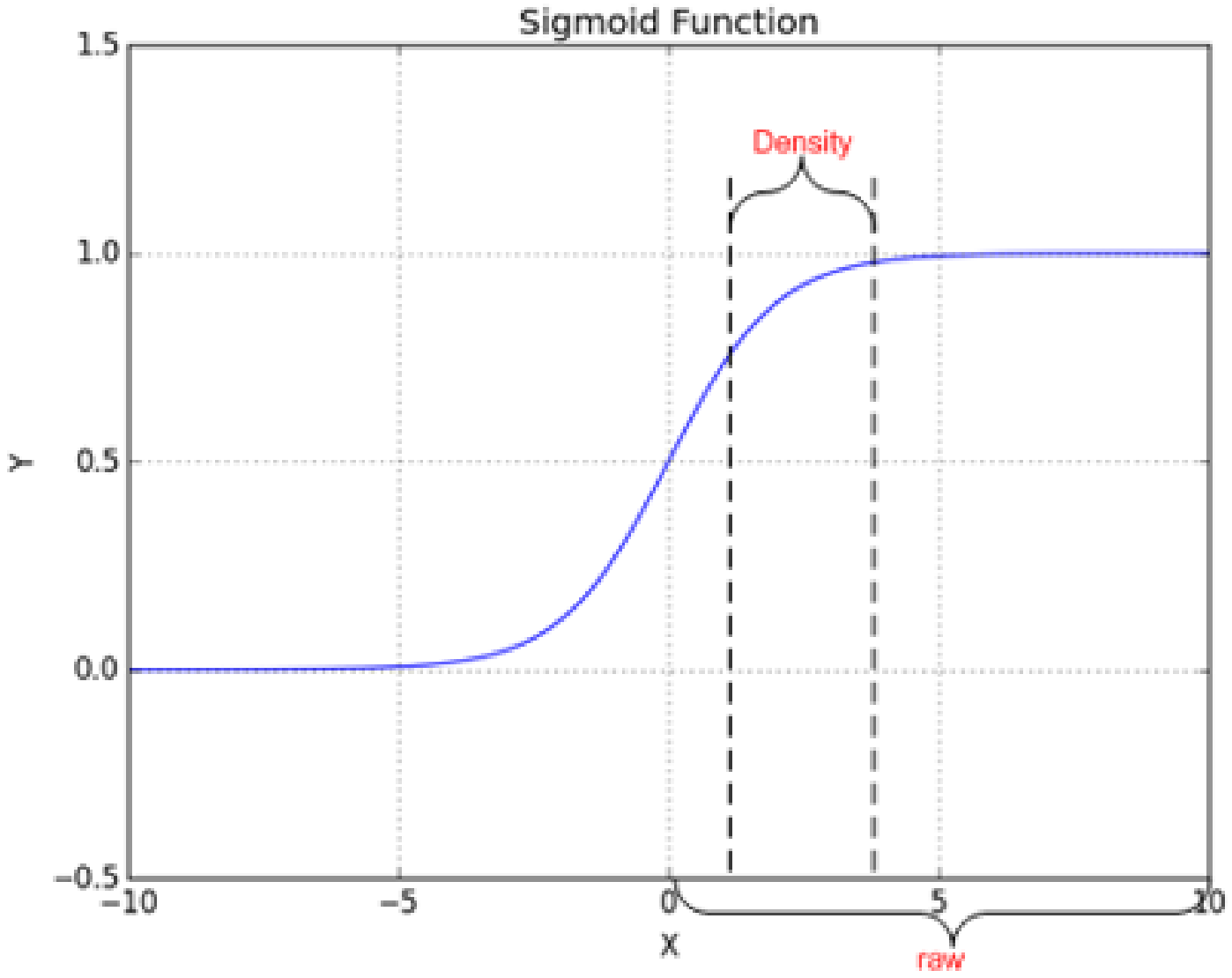} 
  \end{minipage}
  \caption{NVDLA two-stage lookup tables. Left: LRN. Right: Sigmoid.}
  \label{fig:fig8}
\end{figure}

Sigmoid and tanh are more complicated non-linear functions. Since these two functions are both single-valued, they can be calculated by lookup table. Figure \ref{fig:fig7} and \ref{fig:fig8} \cite{nvdla} show two step lookup tables with interpolation. LUT precision is determined by the total lookup points and the slope of the function. The more lookup points, the higher the lookup precision. The steeper the function, the harder the LUT implementation is. Since the hardware resource is restricted, the first step is a raw table which covers the entire domain of definition, while the second step is a dense table with a higher accuracy.

\begin{figure}
  \centering
  \begin{minipage}[t]{0.48\textwidth}
  \centering
  \includegraphics[width=0.6\textwidth]{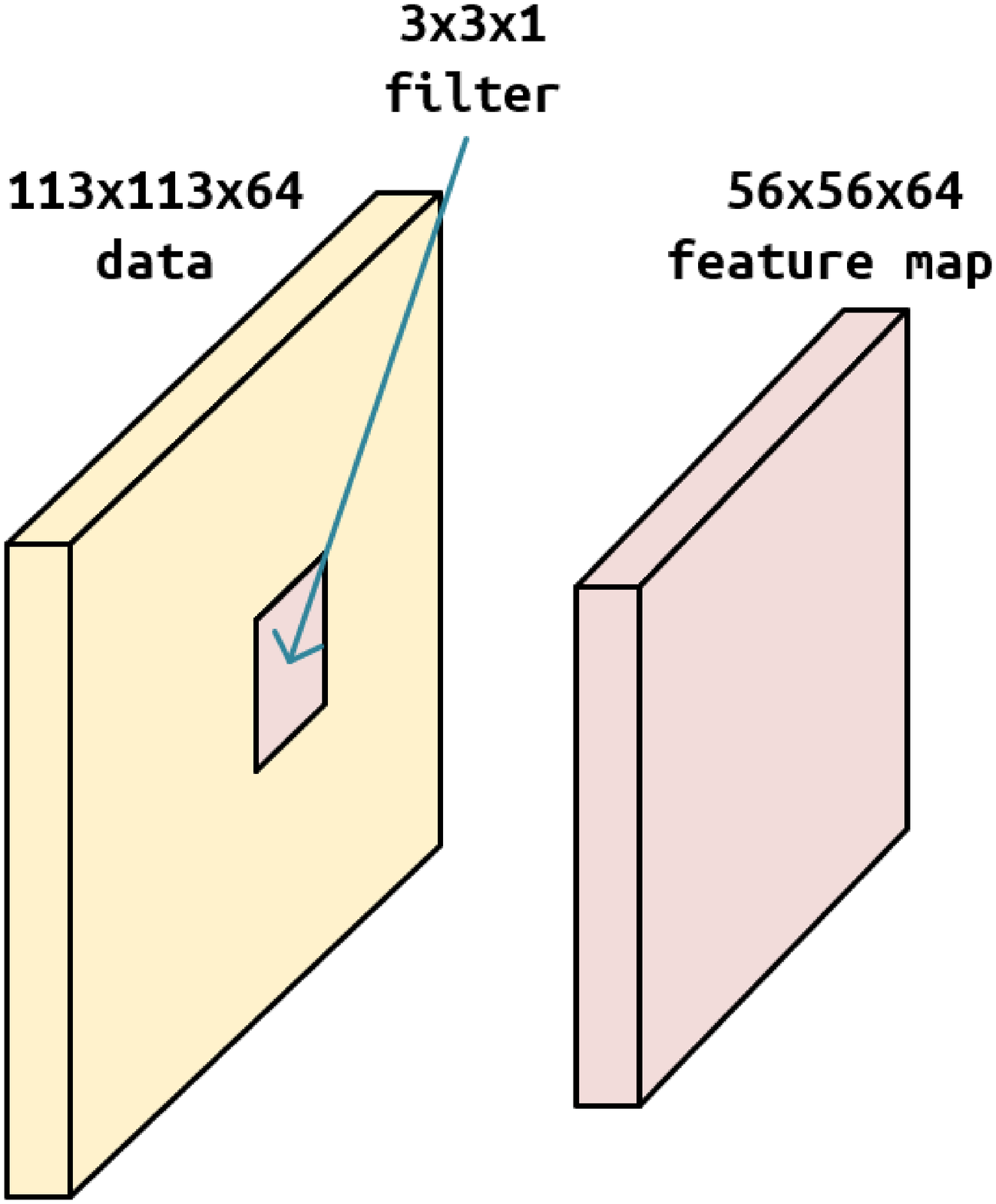} 
  \caption{Pooling diagrams.}
  \label{fig:fig9}
  \end{minipage}
  \begin{minipage}[t]{0.48\textwidth}
  \centering
  \includegraphics[width=0.8\textwidth]{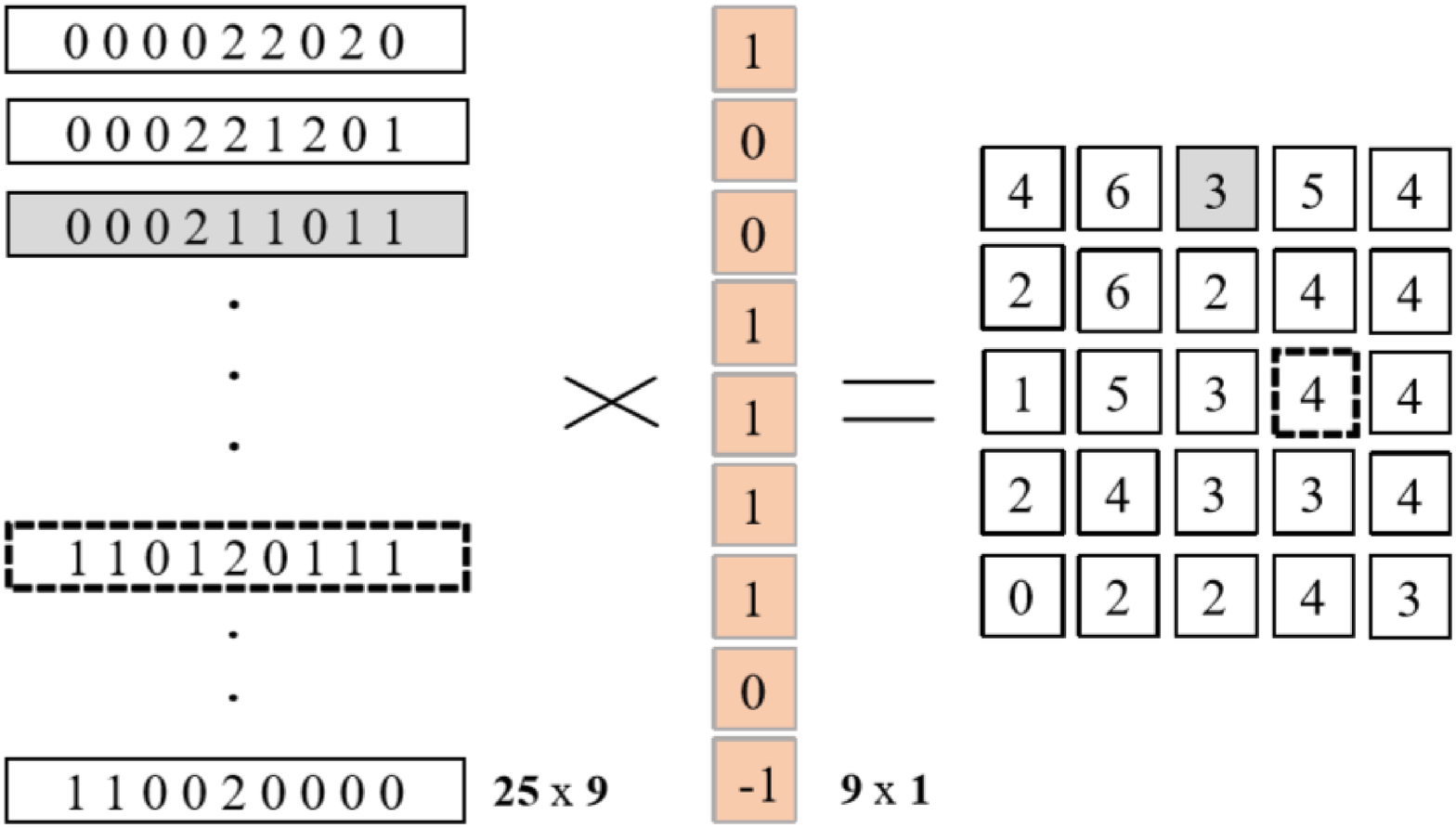} 
  \caption{Im2col convolution process.}
  \label{fig:fig10}
  \end{minipage}
\end{figure}

Figure \ref{fig:fig9} shows pooling operations. The input is normally the output of the previous convolution + activation. Pooling layers serve to preserve the main feature of the previous layer and reduce the data computation during forwarding to prevent over-fitting. Common pooling operations are average-pooling and max-pooling. Average-pooling takes the mean value of the data in the sliding window, while max-pooling takes the max. The output surface of pooling is smaller than the input surface, while the channel dimension remains the same.

\subsection{Optional Optimization Algorithms}
Since LUTs, FFs, DSPs and BRAMs are restricted on FPGA and areas and power consumptions are also concerns on ASIC, it is required to achieve a hardware utilization as high as possible when we choose acceleration algorithms. To deal with memory access and computing in CNN inference, there are algorithms as follows:

\subsubsection{Im2col + GEMM}
Im2col + GEMM is a widely used convolution operation in Caffe. Figure \ref{fig:fig10} \cite{cho2017mec} shows the process. This method transforms the $k\times k\times c$ data matrix and $k\times k\times c$ weight matrix into one-dimension and store them in memories, so that the core can finish the convolution operation by accessing the address one-by-one and doing multiply-accumulate. The matrix multiplication of Numpy library supports BLAS (Basic Linear Algebra Subprograms), which is optimized for linear algebra operations. In Level 3 BLAS, GEMM is supported. GEMM (General Matrix to Matrix Multiplication) is the operation of $\bm{C} =\alpha \bm{AB} +\beta \bm{C}$, which can be directly and efficiently calculated by standard computation libraries.

\subsubsection{MEC (Memory Efficient Convolution)}
\begin{figure}
  \centering
  \includegraphics[width=0.8\textwidth]{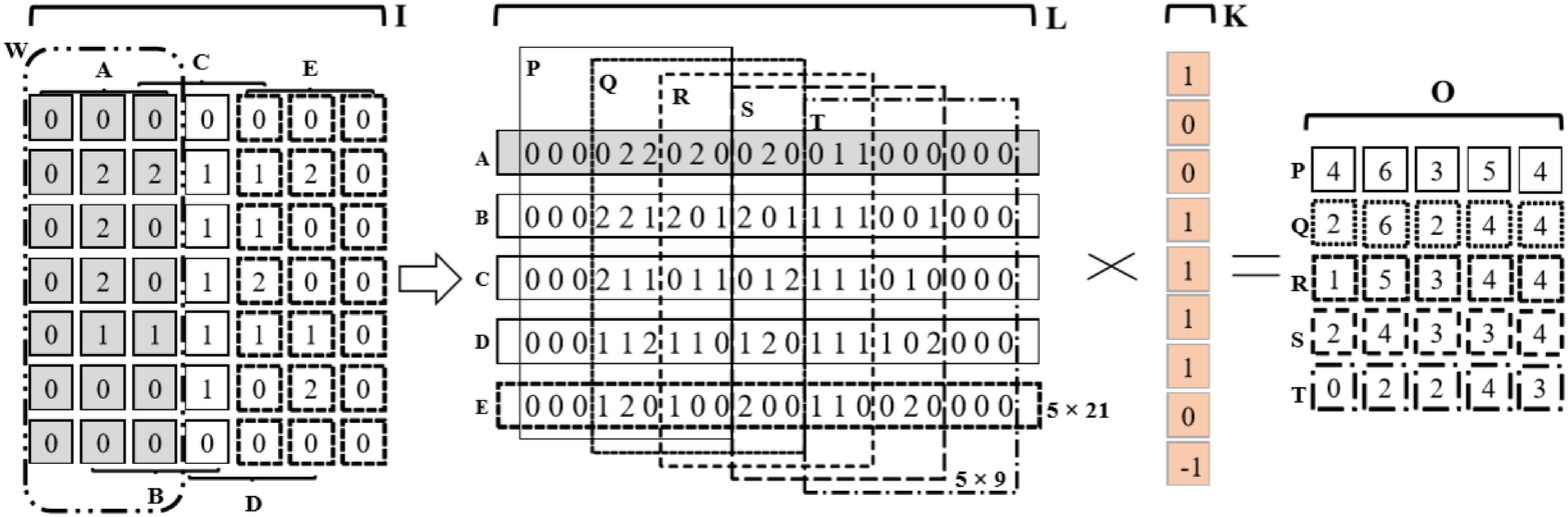} 
  \caption{MEC convolution process.}
  \label{fig:fig11}
\end{figure}

MEC is an extension of im2col + GEMM, as Figure \ref{fig:fig11} \cite{cho2017mec}. Im2col cannot deal with the problem of repeated call of the same data by neighboring matrix multiplication, while MEC uses multiple computation units to solve the problem in pipeline and further decreases the memory access times. It is an essentially trading time for space. For the data section A in Figure \ref{fig:fig11}, the input side is 7, the kernel is 3 and the stride is 1. Since the weight values keep unchanged while sliding on the data section, we can just sequentially reads out \texttt{input\_side * kernel} data, and do pipeline convolution multiply-accumulate operations in a 'STRIDE' of \texttt{stride * kernel}. Such operation generates the first column of output matrix. MEC's advantage is more obvious when there are redundant hardware resources. Its disadvantage is that the scale of pipeline increases with the kernel size, and the computation logic is related to stride. For example, if \texttt{stride} is 2 in Figure \ref{fig:fig11}, matrix multiplication Q and S will be skipped.

\subsubsection{Bitonic Sort}
\begin{figure}
  \centering
  \includegraphics[width=0.3\textwidth]{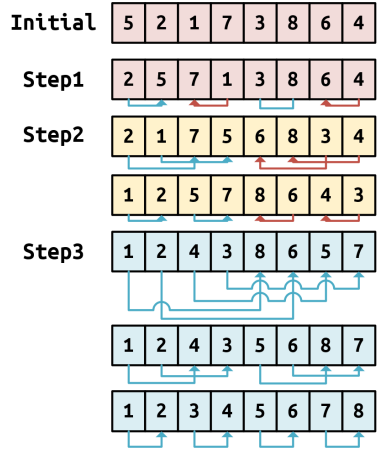} 
  \caption{Bitonic sort example.}
  \label{fig:fig12}
\end{figure}

Bitonic sort is a sorting algorithm towards hardware, as Figure \ref{fig:fig12}. Its main idea is recursively dividing a sequence into two monotonic ascending and descending half-sequences, and the final sequence is sorted. Take the detailed instance of Figure \ref{fig:fig12}: (1) Relating to the second row, sort in ascending and descending order by every two elements. (2) Relating to the third and fourth row, sort in ascending and descending order by every four elements, then by every two elements. After (1) and (2), the first half of the sequence is ascending, and the second half is descending. (3) Relating to the fifth to the eighth row, sort in ascending by every eight elements, then by every four, finally by every two. Then the final sequence is an ascending sequence. Descending sort is the opposite.

Since in every iteration cycle, there are comparison of two numbers in a sequence and its recursive child sequence, the total number of elements must be an integer power of 2 ($n=2^m$). The pros of the algorithm is that $2^{m-1}$ comparators can operate in parallel. Since the total steps are $\log n$ (In Figure \ref{fig:fig12} there are 3 steps), and each step takes maximum $\log n$ group comparison (there are 3 comparisons in the third step in Figure \ref{fig:fig12}), the sequential time complexity of one comparator is $O(n(\log n)^2)$. If $2^{m-1}$ parallel comparators are utilized, the time complexity will be $O((\log n)^2)$. As in Figure \ref{fig:fig12}, 4 comparators can process the 8-number bitonic sort in 6 cycles.

\subsubsection{Pipeline Accumulation}
\begin{figure}
  \centering
  \includegraphics[width=0.8\textwidth]{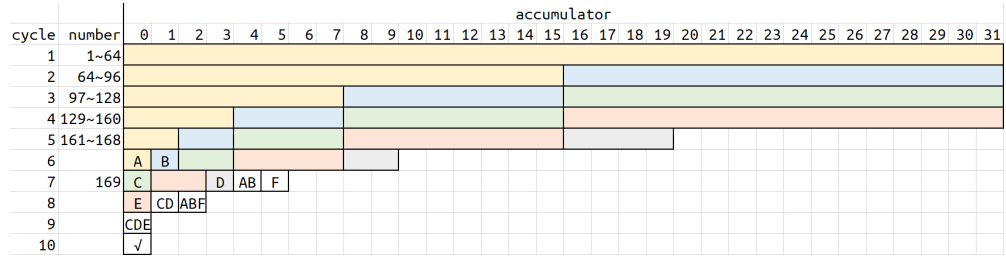} 
  \caption{Pipeline accumulating example.}
  \label{fig:fig13}
\end{figure}

Pipeline accumulation is also an algorithm towards hardware. It is a simple algorithm that get the sum with several adders in groups, trading time for space, as in Figure \ref{fig:fig13}. In Figure \ref{fig:fig13}, 32 adders are calculating the sum of $13\times 13=169$ numbers. In the first cycle, the sum of beginning 64 numbers are calculated. In the second cycle the next 32 numbers are calculated, as well as the partial sum in the previous cycle (the identically-colored part as the previous cycle), and so forth.

The disadvantage of the algorithm is that the most optimized computation unit number is determined by the length of the array. A short of adders in the hardware result in a long calculation time, while a plenty of adder causes some of the adder to be idle after some cycles. Moreover, theoretically we cannot put all adders to operate simultaneously in all cycles, which means there is always a moment that the computation utilization ratio is less or significantly less than 100\%.

Another defect of the algorithm is that if all numbers pending accumulation is all readable, this means that these parallel data must be stored in caches. In the computation process these parallel data are serialized again. Some are accessed in the first cycle, while some are accessed in the following cycles. Such ser-des process decreases the data flow efficiency. On the other hand, if some numbers pending accumulation is readable, the number in every cycle is different in each cycle. In Figure \ref{fig:fig13}, the number of data read out in 10 cycles are 64, 32, 32, 32, 4, 2, 2, 0, 0, 1. Such irregularity would cause the accumulation control logic to be constrained by the convolution kernel size. If there are a variety of convolution kernel sizes in the network, the difficulty to maximize the parallelism would be significant.

In practice, since the width of the memory interface is limited, especially when data volume pending accumulation is large, it is impossible to load out all data in one cycle. As in SqueezeNet v1.1, the kernel size of average-pooling is $14\times 14$ \cite{iandola2016squeezenet} and each number takes 16 bits. If the core reads them all in one cycle, it would take an interface width of $14\times 14\times 16=3136$ bits, which is unrealistic on FPGA. Since memory cannot achieve such bandwidth, the better practice is calculating along with memory readout.

\subsection{Algorithm Trade-off}

\subsubsection{Bitonic Sort \& Pipeline Accumulation}
Whether to use bitonic sort and pipeline accumulation is determined by the data access format in cache. If the dimension of cache is W or H first, then these two algorithms are practical. But if the cache is channel first, it would significantly increase the computation unit number if these two algorithms are utilized (In Figure \ref{fig:fig12}, the computation unit number would be 4 times as comparing one by one). In the final product, such two algorithms are not used.

In this project, the stored data format (NHWC) is optimized for the parallelism of convolution operation, which means the input channel dimension is lowest, while the output channel dimension is the highest (NCHW in Caffe and NHWC in TensorFlow). Such stored data can be directly called as input of the next layer and we do not necessarily have to add extra logic to judge whether to start parallel units or not, as in Figure \ref{fig:fig22}.

\subsubsection{Generic Accelerator vs. Stream Accelerator}
After determining the format of cache, we need to determine whether the cache data is coming from DRAM outside the chip or from the shared DRAM on PC. Since the required on-chip space is different, the corresponding accelerator architecture is different as well. Scalability is another thing to consider when choosing architecture, because the prototype FPGA resource is limited, the maximum parallelism supported is relatively low. If it is deployed on ASIC, there will be much more computation units. A good accelerator architecture decouples from the network, and do not have to change much after being scaled.

\begin{figure}
  \centering
  \includegraphics[width=0.6\textwidth]{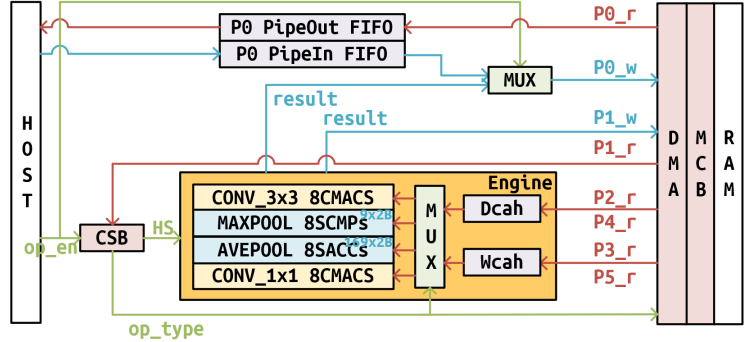} 
  \caption{Generic accelerator architecture.}
  \label{fig:fig14}
\end{figure}

If the required data to calculate come from off-chip DRAMs, the input image, network weight and parameters must be initially loaded to DRAM. After the calculation begins, DMAs moves data from DRAM to cache. Figure \ref{fig:fig14} shows this generic accelerator design. Take the example of the 6 available MCBs (memory controller block) on Spartan-6, two ports can be configured as read/write to be responsible for block data access. The other four ports are read-only ports, which serves to read data and weight (two-way parallel). 6 DMAs serve to generate MCB access timing according to the request from computation unit, control signal block or the host and transfer data between DRAM and these three.

CSB (control signal block) activates the computation core according to the parameters read out from DRAM via DMA and the host commands. In the computation core there are two BRAMs, data cache and weight cache, to cache the block data and weight read out from DMA and provide random access to computation core (since DRAM latency is high). One MUX determines whether data cache and weight cache transfer data to the computation engine or not. P0 write port is shared by two blocks via a MUX. When the system is loading block data, the path from PipeIn FIFO to DMA P0 is selected. When the system is computing, the path from result to DMA P0 is selected.

We need to pay attention that although the block access of DRAM is fast, there is obvious latency. According to the Xilinx datasheet, typical MCB latency of the chip is 22-32 cycles\footnote{\href{https://www.xilinx.com/support/documentation/user_guides/ug388.pdf}{Xilinx UG388 Spartan-6 FPGA Memory Controller User Guide}. Accessed March 6, 2019.}. Since im2col + GEMM operation consists of small pieces of data, the latency due to repeated random DDR accessing would empty the pipeline and waste the parallel computing resource. This part is also discussed in 4.3.3.

\begin{figure}
  \centering
  \includegraphics[width=0.7\textwidth]{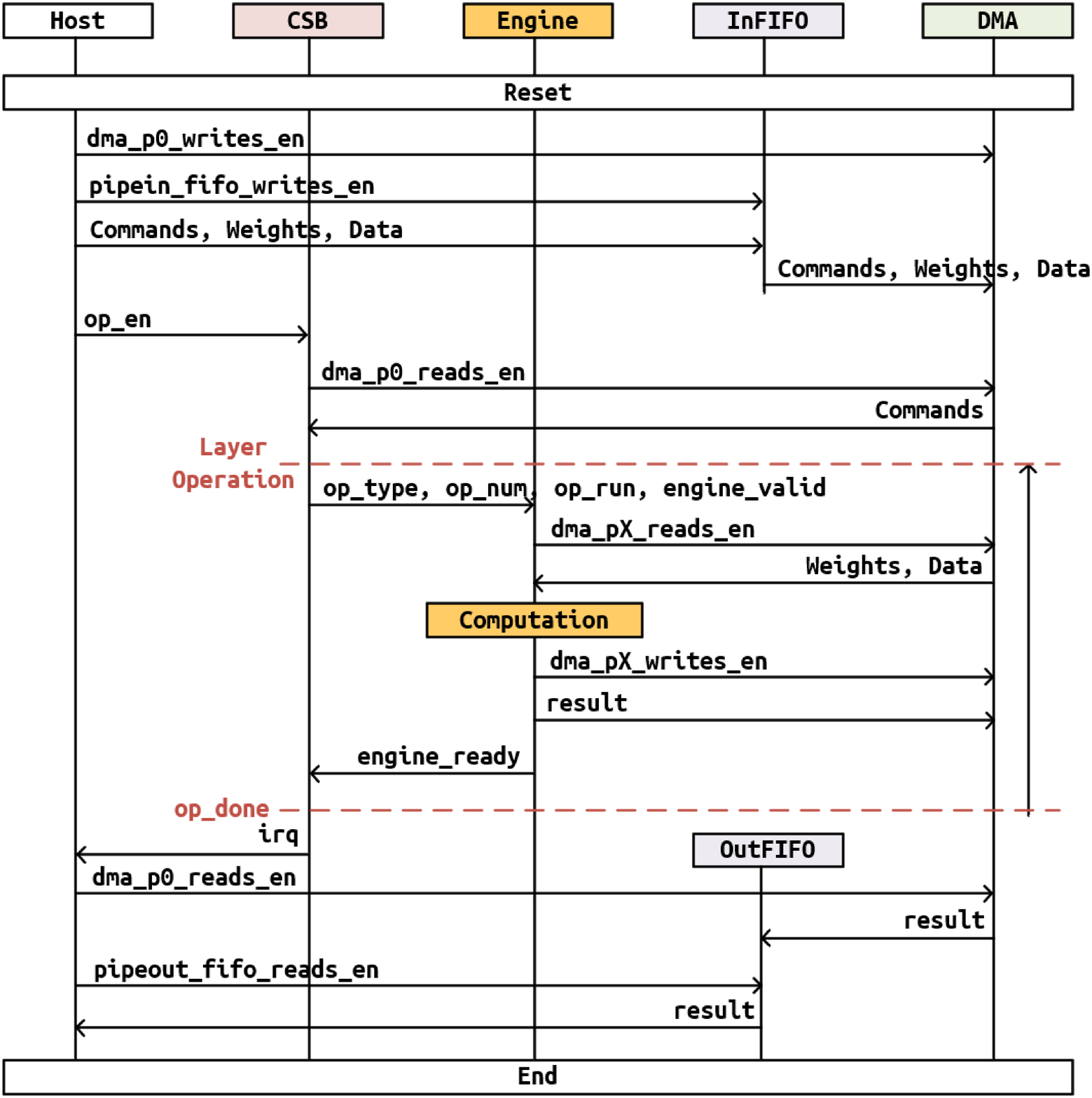} 
  \caption{Operating flow of the generic accelerator.}
  \label{fig:fig15}
\end{figure}

Figure \ref{fig:fig15} shows the flow of generic accelerator. After resetting, the host enables the system. All block data (32 bits x 512, i.e., 512DWORDs) flows from the host to PIPEIN FIFO, then to DRAM. This process includes two clock domains, host clock domain and DRAM clock domain. The write side of PIPEIN FIFO is driven by host clock, while the read side is driven by DRAM clock.

After all data related to inference are loaded to DRAM, the host notifies the control signal block to start the network computation. The computation core extracts the parameter of a layer from DRAM, then reads out data and weight in order. After the computation is completed, the parallel results are serialized and sequentially written back to DRAM. After the writing-back operation is finished, the control signal block sends out an interrupt. Such process loops until all layers are calculated.

The forwarding results are moved from DRAM to PIPEOUT FIFO via DMAs, then to the host via USB3.0. The write side of PIPEOUT FIFO is driven by the DRAM clock, while the read side is driven by the host clock.

To conclude, the entire computation process includes three clock domains, the host clock (100.8MHz), the DRAM clock (333.3MHz) and the computation core clock (100MHz). The entire design hardly meets the timing constraint, takes m logic resources and drains a lot of power after placing and routing.

If we use the generic accelerator, another two complicated problem is on-chip padding and address managing. For the input layers that requires padding, there are two possible solutions. One is to pad before writing back in the previous layer, the other is to pad while reading the data. These two plans lead to different operations at corners.

\begin{figure}
  \centering
  \includegraphics[width=0.8\textwidth]{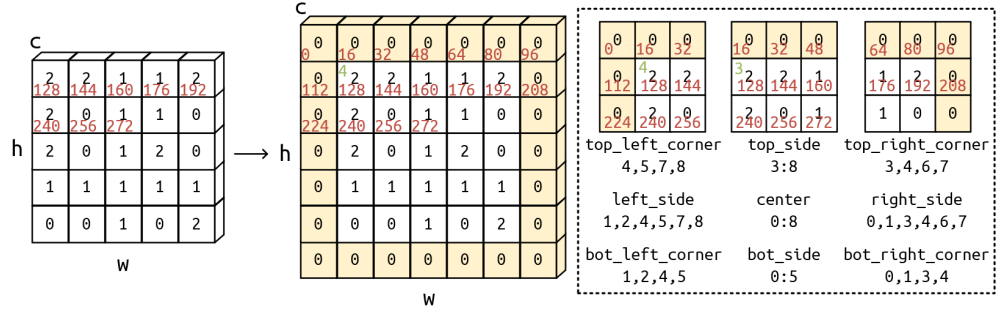} 
  \caption{In-memory padding diagram of generic accelerator.}
  \label{fig:fig16}
\end{figure}

Figure \ref{fig:fig16} shows an example of padding while reading. The parallelism in Figure \ref{fig:fig16} is 16. When this layer is going to write back the $5\times 5\times c$ results on the left side and the next layer requires padding = p, the DMA needs to jump access DRAM to reserve the positions of 0 (in this case start writing back from 128). After finishing the write-back of a row, the DMA should jump \texttt{2p * BURST\_LEN} (channel parallelism). Then in the next layer this result can be accessed from address 0.

If padding before the current layer, the core needs to judge if the current convolution kernel is at the corners or sides, which leads to 9 situations in Figure \ref{fig:fig16}. When the convolution kernel is at the top left corner, the core just reads address 4, 5, 7 and 8. When the convolution kernel is at the top side except corners, the core reads address 3-8. Such design strategy would bring about many judgment statements and dependencies, which increase LUT resource utilization after synthesis.

No matter which padding strategy is taken, in terms of address management, the result block written back to DRAM should have the same dimension order as the input block read out, so that coherency of the flow is preserved. Since the input data cube is stored in NHWC (input channel as the lowest dimension) in which the result channel of the convolution is at the highest dimension, memory reshaping is required to adjust the output matrix dimension. This part would be complicated. Meanwhile, when a convolution core is calling a block of data, suppose that the order is NHWC rather than NWHC, after \texttt{kernel} pieces of data are read out in the row direction, the DMA needs to jump to the next row. The jump length is \texttt{BURST\_LEN * (input\_side - kernel)}. After calculating the first column of output, the DMA jumps back to the second input column. Such logic would introduce a lot of registers to calculate the shift value. Moreover, outputs of parallel convolution layers (e.g., expand1x1 and expand3x3 in SqueezeNet v1.1) are in NWHC, while the concatenation layer merges these two matrices in the lowest channel dimension. This requires some extra logic as well.

If the data are directly from the host, we just need to transfer every piece of cache data to FPGA via high-speed interface rather than storing it on extra off-chip DDR. This is a stream architecture that shares memory with the host, like NVDLA. However, USB3.0 high-speed IO has latency as well. The total IO operation latency is USB latency + OS latency + storage latency.

The final design architecture is tha latter. The main reason is that if the data are initially moved to off-chip DDR, the complex logic from DDR to cache must be redesigned as well. On the other hand, such operations are mature on X86/ARM CPUs with C++/Numpy libraries.

\subsubsection{Im2Col vs. MEC - Channel-first Parallelism vs. Surface-first Parallelism}
For operations like convolution and pooling, there are two optional parallel solutions. Channel-first parallelism means parallel in the channel dimension. Multiple multiply-accumulators calculates the parallel channel data, followed by multiple channel accumulators. The pros are that in all cycles the parallelism is the highest while the cons are that one piece of data is called for multiple times in neighboring convolutions. If the access speed of cache is faster than calculation speed, such method would not increase halt time in the computation pipeline.

Surface-first parallelism means parallel in H or W dimension of the input matrix to address the shared neighboring problem, which is the idea of MEC. When the convolution starts, the parallel computation units are not all activated. The parallelism is highest when the convolution goes to the intermediate positions, as in Figure \ref{fig:fig19} and Figure \ref{fig:fig20}. The advantage of this solution is that every single data is accessed from cache for once and shared by multiple weights, so that the data memory is accessed less.

The cons of this solution are that in the whole operation process, the computation parallelism varies, and the logic is pretty complicated. We need to adjust and prepare different parallel slots for different strides, since stride determines the max parallelism of a convolution. For every two neighbor convolution, there are \texttt{kernel * (kernel - stride)} numbers that are overlapped. Notice that the multiplied weights to these overlapped data in two convolutions are different. So we need multiple groups \texttt{kernel - stride + 1} of parallel computation units to process that. When \texttt{kernel} is 3 and \texttt{stride} is 1, three slots will be occupied (In Figure \ref{fig:fig19}, \texttt{sum\_enable = 111}) and the data in the middle will be accessed for three times in three convolutions. When \texttt{kernel} is 3 and \texttt{stride} is 2, there is a slot that is always empty, and the data in the middle will be accessed for only two times.

Moreover, if \texttt{kernel} increases (e.g., in AlexNet there is \texttt{kernel size} of $11\times 11$), the slot number required increases proportionally. It is not a good practice since it makes the hardware size constrained by network size. Unless the slot size is very huge, networks with large convolution kernels are not supported. It is neither a runtime configurable design anyway.

\begin{figure}
  \centering
  \includegraphics[width=0.8\textwidth]{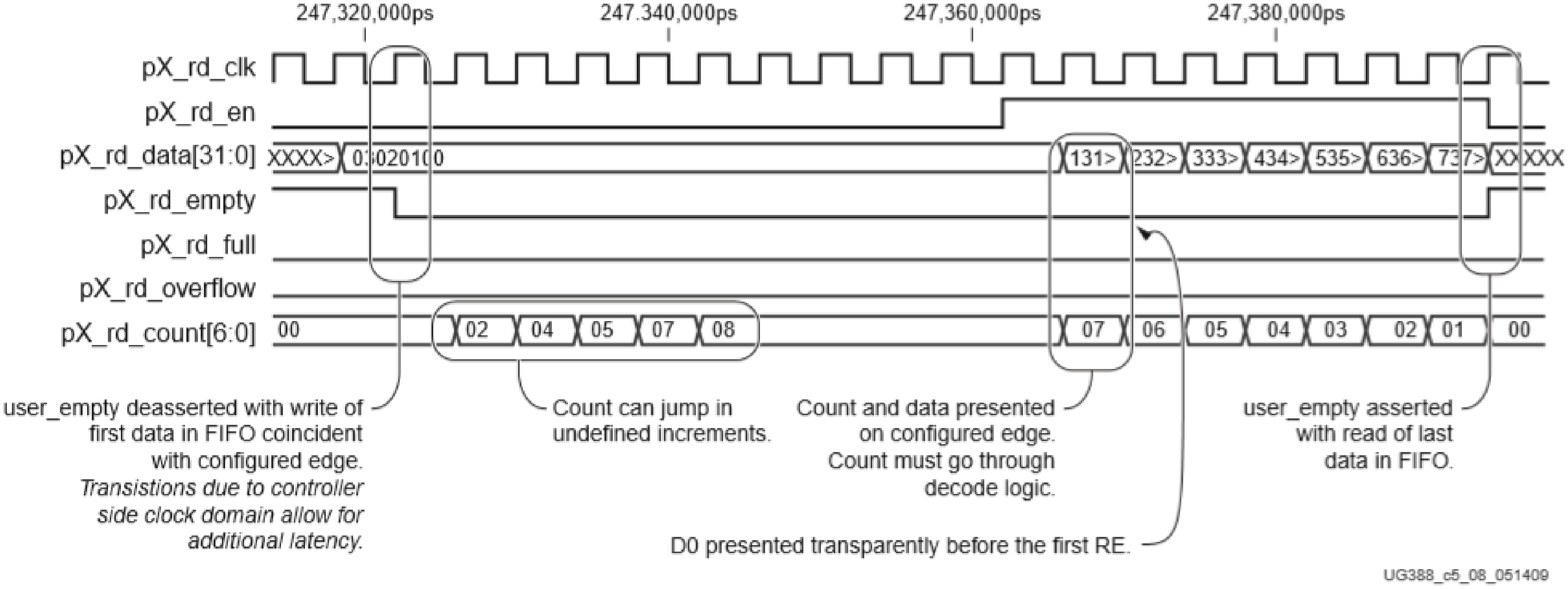} 
  \caption{Xilinx MCB read timing diagram.}
  \label{fig:fig17}
\end{figure}

\begin{figure}
  \centering
  \includegraphics[width=0.7\textwidth]{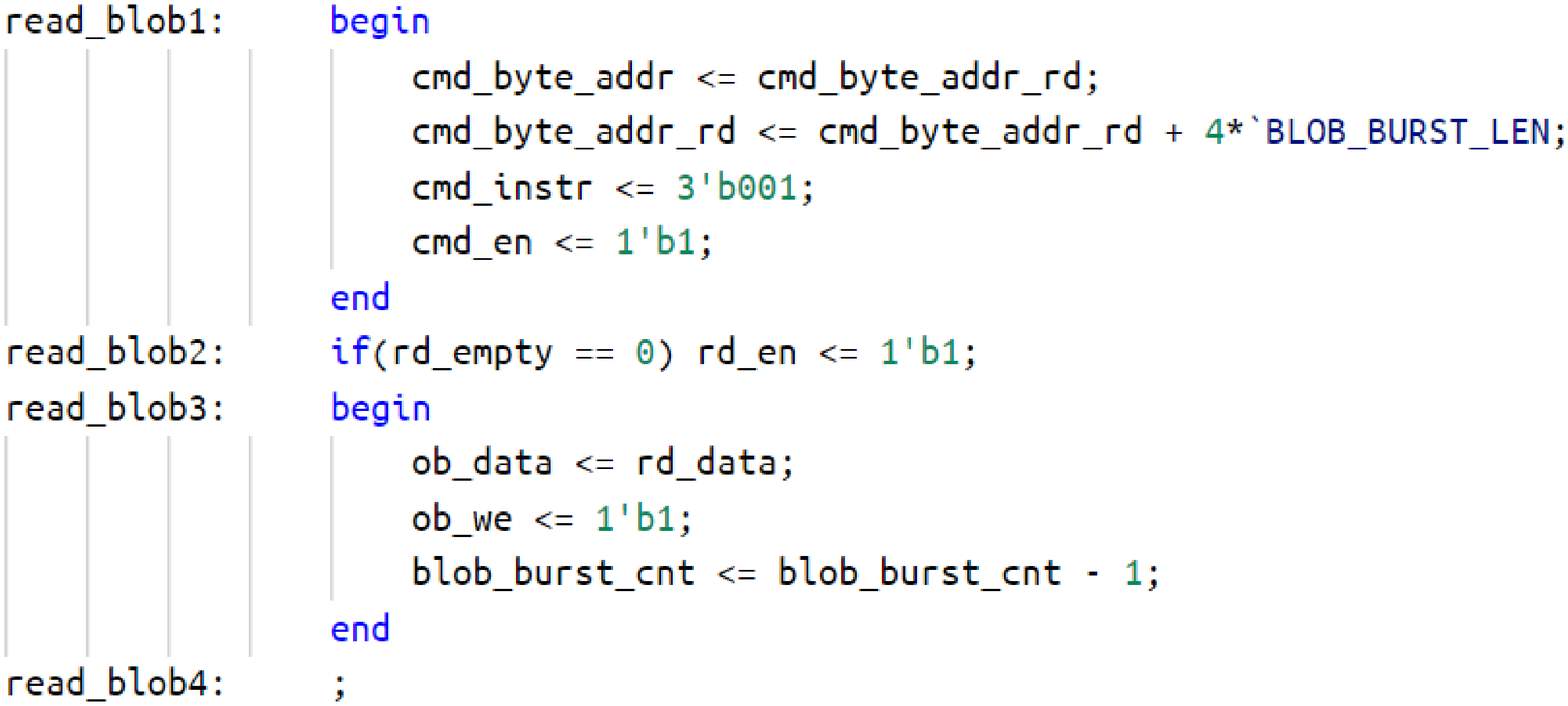} 
  \caption{MCB read timing and DMA code.}
  \label{fig:fig18}
\end{figure}

\begin{figure}
  \centering
  \includegraphics[width=1.0\textwidth]{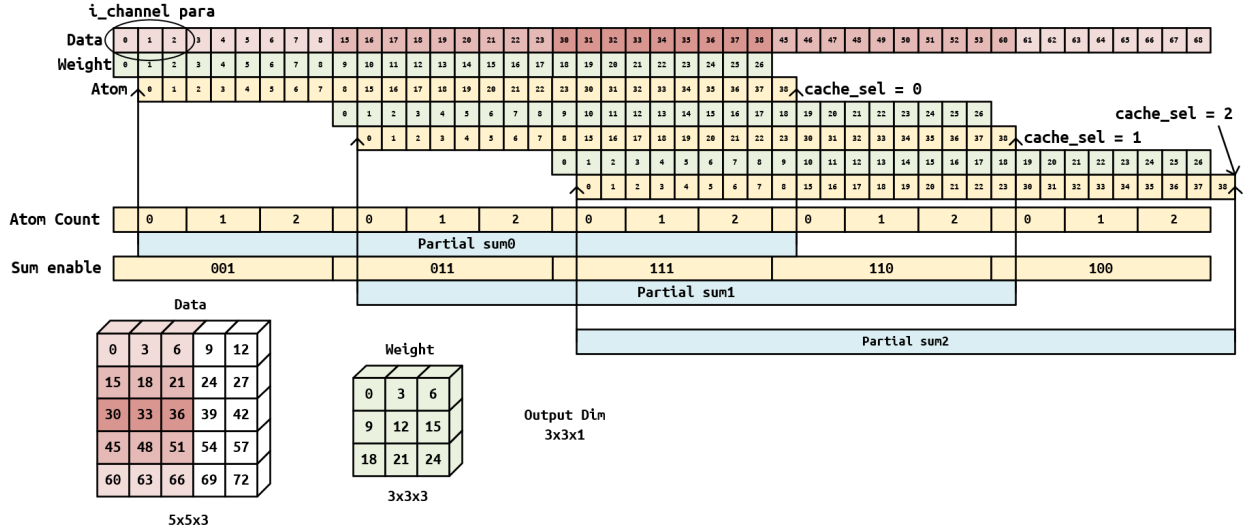} 
  \caption{MEC convolution diagram (stride = 1, input channel size = 3).}
  \label{fig:fig19}
\end{figure}

\begin{figure}
  \centering
  \includegraphics[width=1.0\textwidth]{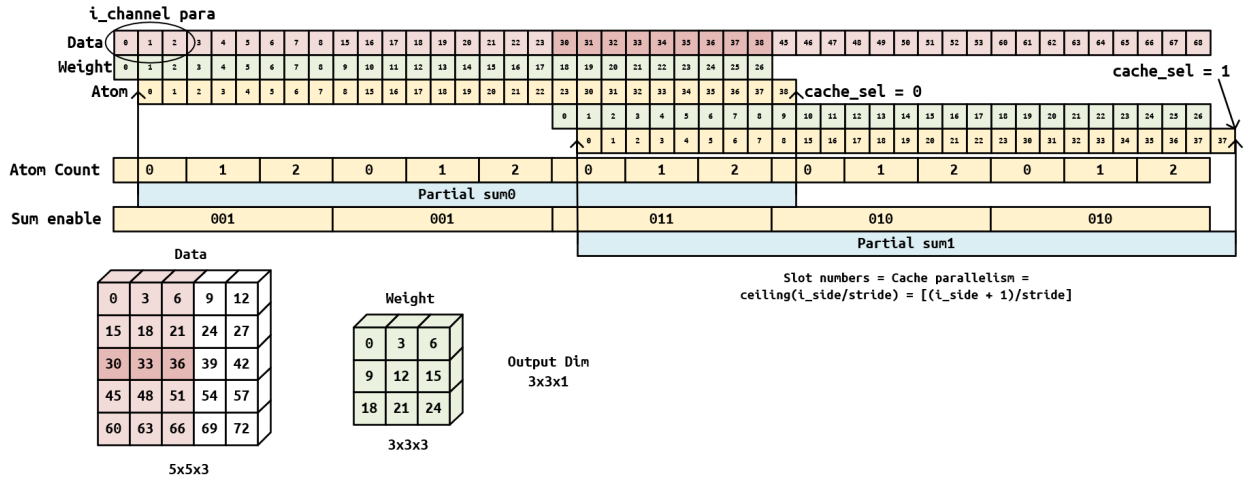} 
  \caption{MEC convolution diagram (stride = 2, input channel size = 3).}
  \label{fig:fig20}
\end{figure}

This project uses channel-first parallelism. Because the data are cached in BRAM that requires only one cycle for each readout, which is significantly faster than computation units. No empty pipelines or idle computation units would exist in the whole flow and it leads to a simpler design. If the data are stored in DRAM, we can refer to Xilinx MCB IP readout sequence in Figure \ref{fig:fig17}. DMA takes at least 4 cycles to readout data from Xilinx MCB in Figure \ref{fig:fig18} \footnote{\href{https://www.xilinx.com/support/documentation/user_guides/ug388.pdf}{Xilinx UG388 Spartan-6 FPGA Memory Controller User Guide}. Accessed March 6, 2019.}. In the first cycle the DMA sends a read access command to MCB. In the second cycle the MCB read enable \texttt{pX\_rd\_en} is pulled high after judging the MCB IO type. In the third cycle data is read out (if there is only one). The fourth cycle is idle.

Another reason channel-first parallelism is chosen is that channel dimensions of convolution networks are normally integer times of 4, 8 or 16 except the initial input image. Such feature makes computation units to scale more easily, we do not need to consider the problem of padding 0 in the input channel dimension except the initial layer whose channel is 3. Contrarily, the sizes of intermediate layer surfaces vary in different networks, it is difficult to schedule in the surface-first parallelism solution.

\section{Hardware Architecture}
FP16 formats are used in the storage and computation of the final design, because FP16 models do not have to be quantized and retrained from FP32 like INT8 and FP16 models saves 50\% storage space and computation units compared to FP32 models. On the other hand, the activation layers and the softmax operation at the end make the forwarding process not sensitive to the deviation between FP16 and FP32, so FP16 do not lose much accuracy in inference. FP16 ranges from $6\times 10^{-5}$ to $6\times 10^4$ and the precision is 0.05\%, while FP32 ranges from $1\times 10^{-38}$ to $1\times 10^{38}$ and the precision is 0.000006\% \cite{han2017efficient}. Since the three channels of the input image are remapped to [0, 255] and followed by mean processing, the range of FP16 is enough for intermediate computations.

\begin{figure}
  \centering
  \includegraphics[width=0.3\textwidth]{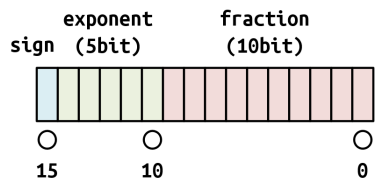} 
  \caption{FP16 data format.}
  \label{fig:fig21}
\end{figure}

\begin{figure}
  \centering
  \includegraphics[width=0.8\textwidth]{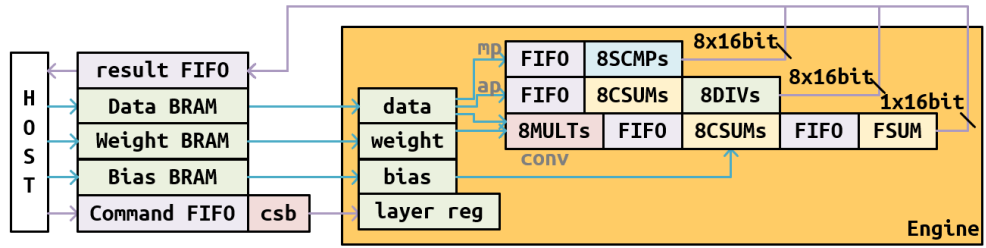} 
  \caption{Stream Accelerator architecture.}
  \label{fig:fig22}
\end{figure}

\begin{figure}
  \centering
  \includegraphics[width=0.7\textwidth]{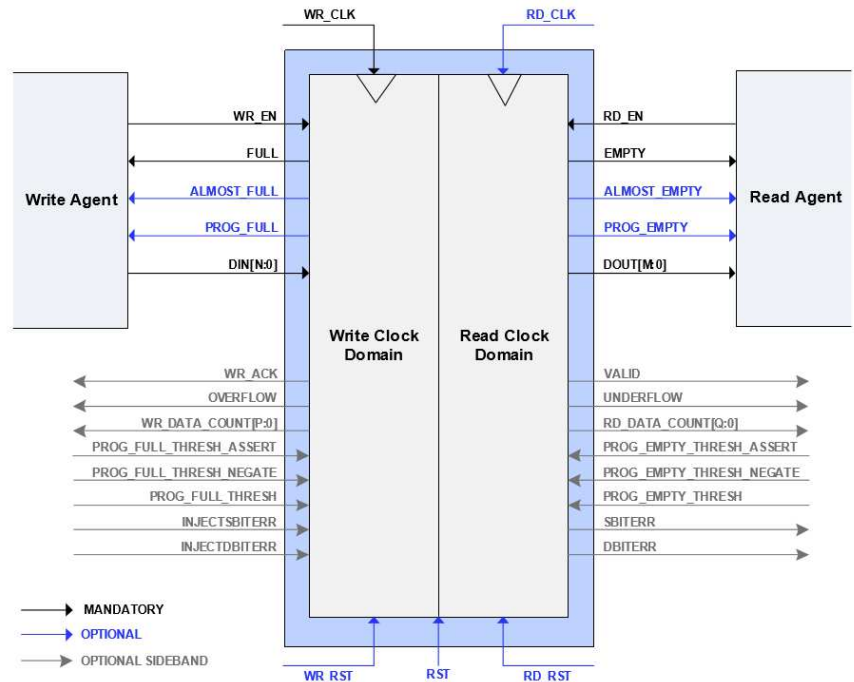} 
  \caption{FIFO with independent clock domains.}
  \label{fig:fig23}
\end{figure}

Figure \ref{fig:fig22} shows the stream accelerator architecture consisting of the following blocks: control signal block, engine, usb communication block (Host), several FIFOs and BRAMs. All FIFOs in the design are asynchronous FIFOs with handshake, supporting independent read/write clock domains, as in Figure \ref{fig:fig23} \footnote{\href{https://www.xilinx.com/support/documentation/ip_documentation/fifo_generator/v13_1/pg057-fifo-generator.pdf}{Xilinx PG057 FIFO Generator v13.1 LogiCORE IP Product Guide}. Accessed March 6, 2019.}. For command FIFO, the write clock domain is the USB clock while the read clock domain is the engine clock. For result FIFO, the write clock domain is the engine clock while the read clock domain is the USB clock.

The engine module consists of registers and computation units. Registers are data, weight, bias and layer registers. Computation units consist of several parallel floating-point computation modules and FIFOs. FIFOs serve to deal with the speed matching of different types of floating-point computations.

\subsection{Control Signal Block and Parameters}
Control signal block parses the required parameters for every layer of the network from the USB3.0 packets and store them in to layer registers.In the computation process these registers are accessed. Table \ref{tab:table1} \footnote{\href{https://resources.wolframcloud.com/NeuralNetRepository/resources/SqueezeNet-V1.1Trained-on-ImageNet-Competition-Data}{Wolfram Squeezenet v1.1 Trained on Imagenet Competition Data}. Accessed March 6, 2019.} shows the network tuple of SqueezeNet v1.1 and the detailed parameters are in Table \ref{tab:table2}. SqueezeNet v1.1 consists of four types of calculation, convolution $1\times 1\times c$, convolution $3\times 3\times c$, max-pooling $3\times 3$ and average pooling $14\times 14$. ReLu operations can be directly realized after convolution. Concatenation layers can be realized by Numpy matrix operations. Since SqueezeNet v1.1 uses a lot of $1\times 1$ convolution kernels and the strategy of expanding after squeezing. The maximum convolution core is 3x3 and the network weights are reduced by 50 times compared to AlexNet while the inference accuracy is relatively the same\cite{iandola2016squeezenet}.

\begin{table}
    \centering
    \caption{SqueezeNet v1.1 network structure.}
    \begin{tabular}{l|l|l}
    Name        & Type & Dimension \\
    \hline
    input       & Input & 3-tensor (size: 3x227x227) \\
    conv1       & Convolution Layer & 3-tensor (size: 64x113x113) \\
    relu\_conv1  & Ramp              & 3-tensor (size: 64x113x113) \\
    pool1       & Pooling Layer     & 3-tensor (size: 64x56x56) \\
    fire2       & Net Graph (7 nodes) & 3-tensor (size: 128x56x56) \\
    fire3       & Net Graph (7 nodes) & 3-tensor (size: 128x56x56) \\
    pool3\_pad   & Padding Layer     & 3-tensor (size: 128x57x57) \\
    pool3       & Pooling           & 3-tensor (size: 128x28x28) \\
    fire4       & Net Graph (7 nodes) & 3-tensor (size: 256x28x28) \\
    fire5       & Net Graph (7 nodes) & 3-tensor (size: 256x28x28) \\
    pool5\_pad   & Padding Layer     & 3-tensor (size: 256x29x29) \\
    pool5       & Pooling Layer     & 3-tensor (size: 256x14x14) \\
    fire6       & Net Graph (7 nodes) & 3-tensor (size: 384x14x14) \\
    fire7       & Net Graph (7 nodes) & 3-tensor (size: 384x14x14) \\
    fire8       & Net Graph (7 nodes) & 3-tensor (size: 512x14x14) \\
    fire9       & Net Graph (7 nodes) & 3-tensor (size: 512x14x14) \\
    drop9       & Dropout Layer     & 3-tensor (size: 512x14x14) \\
    conv10      & Convolution Layer & 3-tensor (size: 1000x14x14) \\
    relu\_conv10 & Ramp              & 3-tensor (size: 1000x14x14) \\
    pool10      & Aggregation Layer & vector (size: 1000) \\
    flatten     & Flatten Layer     & vector (size: 1000) \\
    probabilities & Softmax Layer   & vector (size: 1000) \\
                & Output            & class
    \end{tabular}
    \label{tab:table1}
\end{table}

\begin{table}
  \centering
  \caption{SqueezeNet v1.1 network parameters and the configuration commands required by this project.}
  \includegraphics[width=1.0\textwidth]{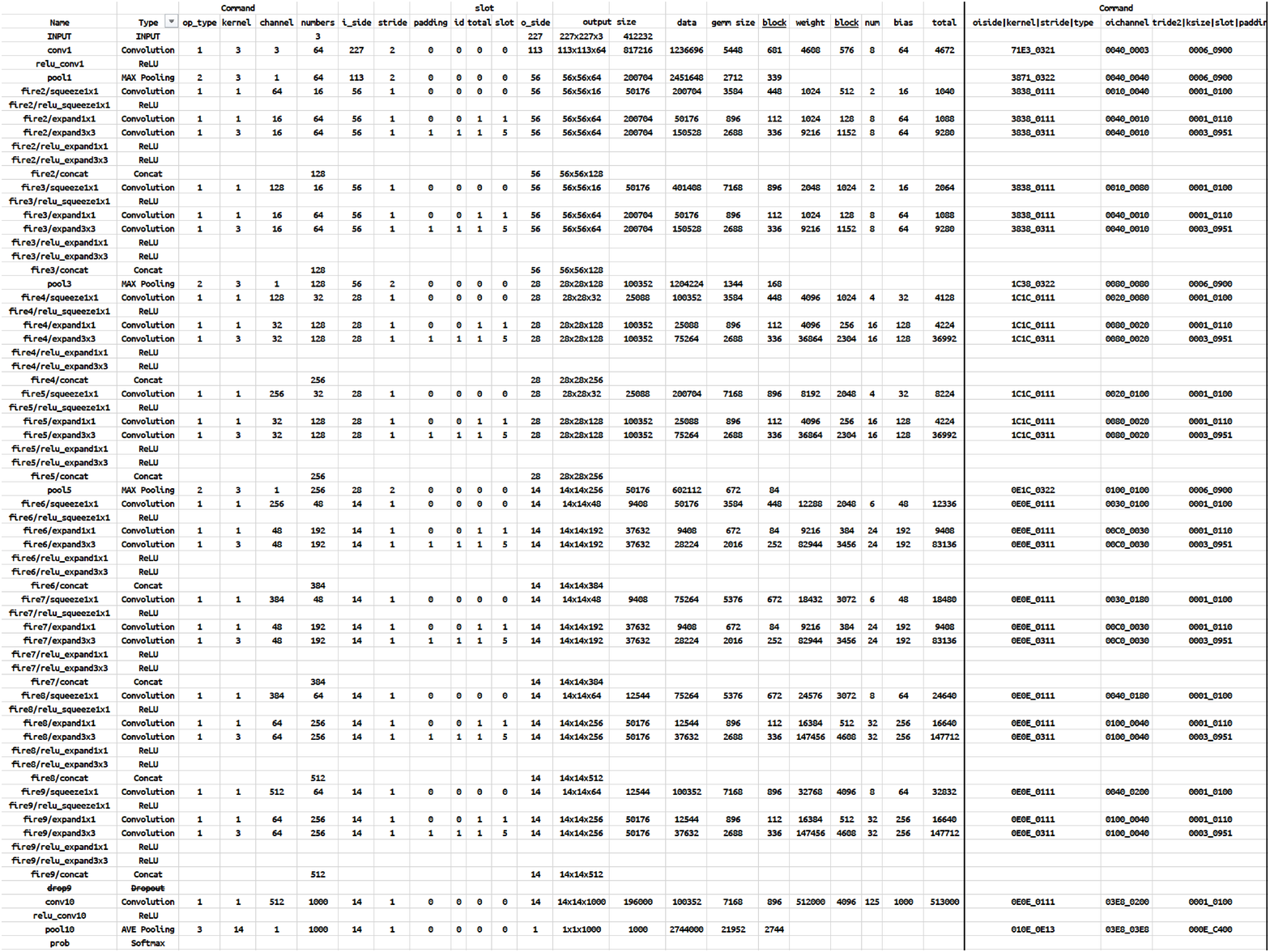} 
  \label{tab:table2}
\end{table}

\subsection{Engine}
Engine computes according to the layer register. The operating flow is as Figure \ref{fig:fig36}. There are three registers, data, weight and bias in the computation units (parallelism = 8) in Figure \ref{fig:fig22}, whose widths are 128 bits, 128 bits and 16 bits independently. Layer registers store the parsed parameters of a single layer (12bytes, as in Figure \ref{fig:fig34}). Computation units consist of three sections, convolution, max-pooling and average-pooling. Each section consists of several floating-point units and FIFOs. The parallelisms are all 8.

At 100MHz clock speed, FP16 multiplier latency is 6 cycles, FP16 adder latency is 2 cycles, FP16 comparator latency is 2 cycles and FP16 divider latency is 6 cycles. Since the adders in the computation units are used as accumulators, and the comparators are also continuously operating, new data should be fed after the accumulators or comparators are finished rather than in every cycle. Multipliers can be fed in every cycles as in MULT in Figure \ref{fig:fig26}. Hence, the throughput of multipliers is larger than that of accumulators or comparators. FIFOs are required to match the input data update frequency of accumulators and comparators.

\subsubsection{Convolution Units}
Convolution units consist of 8 parallel floating-point multipliers, 8 parallel floating-point adders, 2 FIFOs and an independent floating-point adder. The calculation formula is as follows, in which $D$ stands for the elements of the input matrix (padded), $W$ stands for the weights in convolution kernels, $A$ stands for the elements in the output matrix, $wo$, $ho$, $co$, $n$ stands for the output dimensions, and $w$, $h$, $ci$ stands for the input dimensions, and similarly hereinafter.

\begin{equation}
    A _{wo,ho,co,n} = B _{n} + \sum _{c=0}^{c=ci} \sum _{h=ho\cdot s}^{h=ho\cdot s + k} \sum _{w=wo\cdot s}^{w=wo\cdot s + k} W _{w,h,c,n} \times D _{w,h,c}
\end{equation}

\begin{figure}
  \centering
  \includegraphics[width=0.8\textwidth]{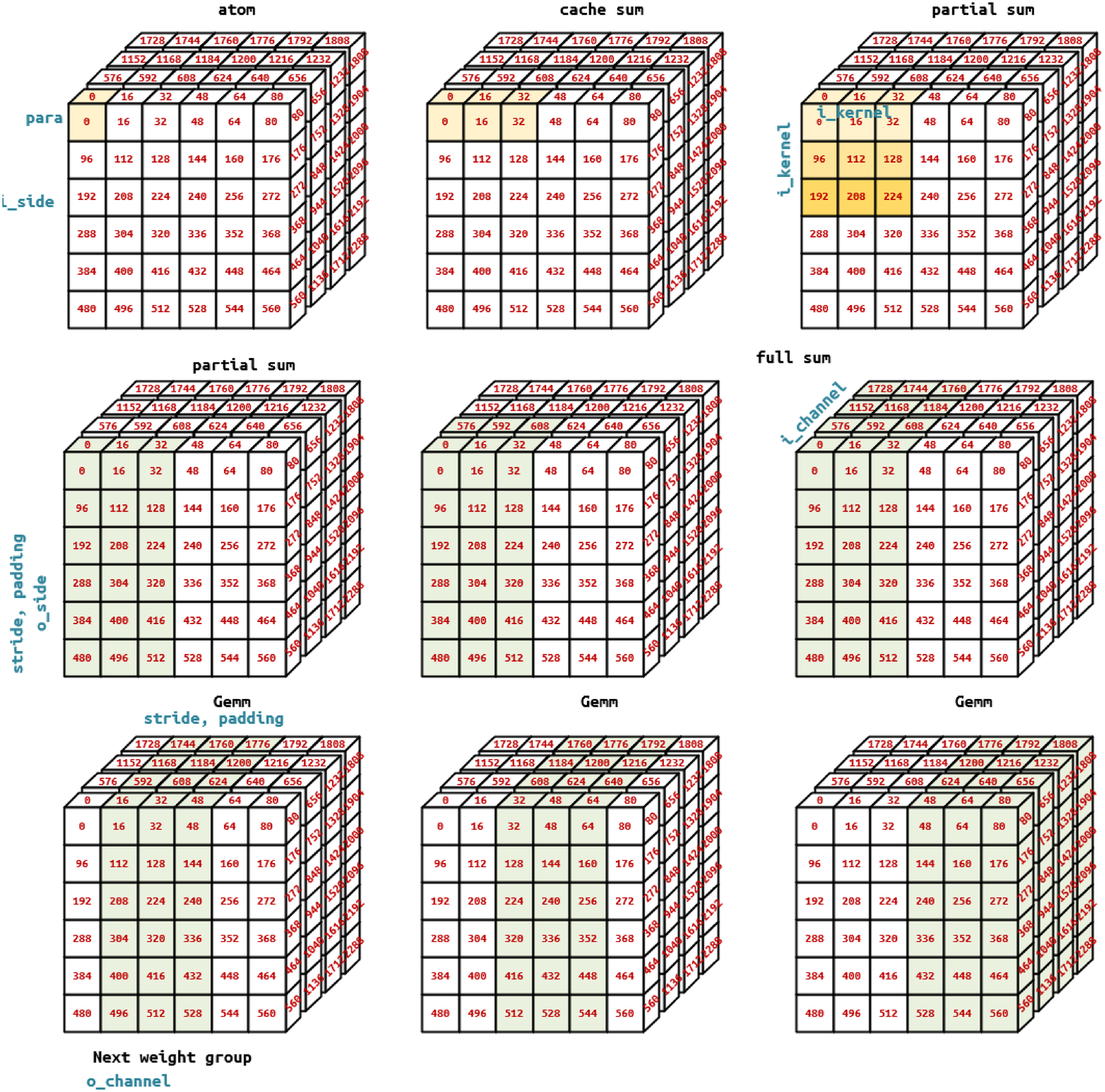} 
  \caption{Convolution process (parallelism = 16).}
  \label{fig:fig25}
\end{figure}

Figure \ref{fig:fig25} shows the macroscopic im2col convolution (input data access only). The parallelism is 16. The convolution process goes in five dimensions from the lowest to the highest. The operation in the lowest dimension is $data\times weight$. It is the smallest operation in the while process, called as \emph{atom}. The second dimension is the filter movement in H or W direction after calculating the partial sum of a filter kernel. The third dimension is the traversal of the entire input channel multiply-accumulation of one column or row. The fourth dimension is the filter movement of a stride in W or H direction after channel operations, until the whole surface is finished. The final dimension is the traversal of all $n$ output channels, which means calling all $n$ filters to do convolution operation to the same input matrix.

\begin{figure}
  \centering
  \includegraphics[width=1.0\textwidth]{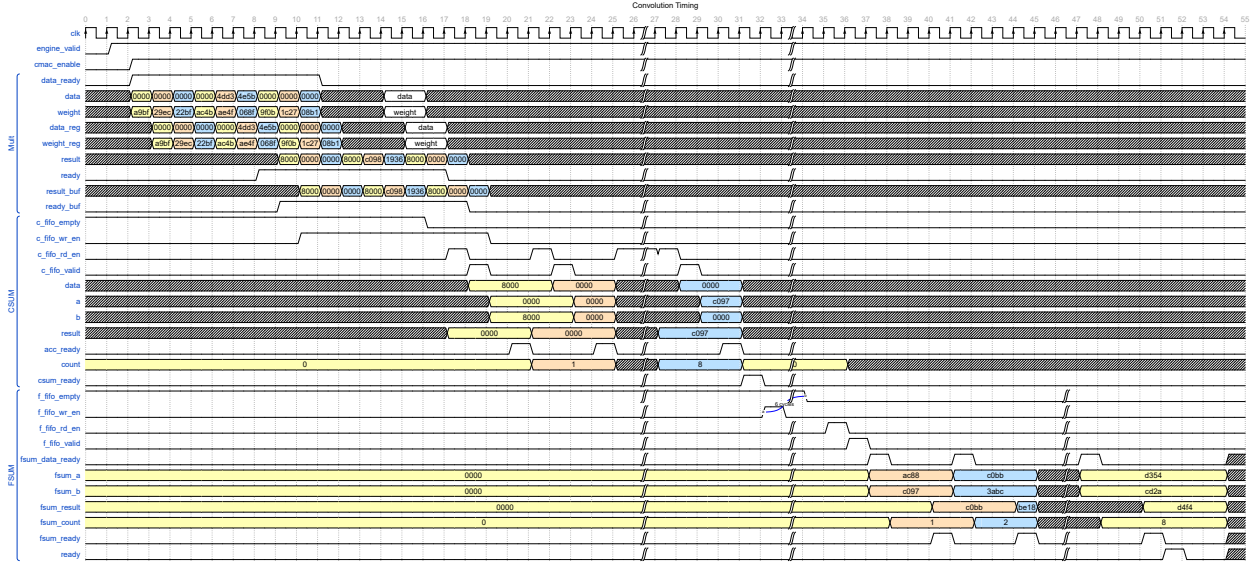} 
  \caption{Convolution timing sequence.}
  \label{fig:fig26}
\end{figure}

Figure \ref{fig:fig26} shows the microscopic RTL convolution process. The parallelism is 8 and it is a three-stage pipeline flow. After \texttt{engine\_valid} is high, \texttt{cmac\_enable} is pulled high and computation units are enabled. The computation units read out data, weights and biases from BRAM (the bias and internal transfer signals are not shown). Notice that when \texttt{data\_ready} is low, the values fed to floating-point computation units will not be calculated, and so forth. When \texttt{data\_ready} is high, data and biases are continuously fed to 8 parallel multipliers (Figure \ref{fig:fig26} shows only 1). After 6 cycles (In Figure \ref{fig:fig26}, \texttt{kernel size} is $3 \times 3 = 9$) continuous results are calculated. After calculation, \texttt{ready} is pulled high, and the results are stored into partial sum FIFO. Write latency is 6 cycles, then \texttt{p\_fifo\_empty} is pulled low.

After the multiply calculation, 8 parallel partial sum accumulators start. Each one reads out \texttt{kernel} pieces of data from the \texttt{P\_FIFO}, and do accumulations. After PSUM accumulation, \texttt{psum\_ready} is pulled high, and the results are stored in full sum FIFO. Write latency is 6 cycles, then \texttt{f\_fifo\_empty} is pulled low.

After PSUM calculation, the final-stage single full sum accumulator sums the 8 partial results in the previous stage. Notice that the initial value is the bias (0xac88 in Figure \ref{fig:fig26}). After the calculation \texttt{fsum\_ready} is pulled high and the data is stored in \texttt{fsum}. \texttt{fsum} is 16 bits wide. Since the first feature map surface sizes of networks based on ImageNet after the first convolution are smaller than 128, \texttt{fsum} depth is set at 128. This cache is inferred as a single-port RAM by ISE. The initial value in \texttt{fsum} cache in future steps are the previous accumulating results of FSUM process.

Notice that there are two FIFOs in the three-stage pipeline to match the different speed of three computation modules. They also help to adjust the computation speed of each floating-point IP during design. Because the lower the latency is, the more combinational logic resources are taken, the more difficult it is for the timing to converge, and the higher the fan-out is. The FIFOs do not affect timing but help decouple the stages. In ASIC projects there are floating-point IPs of 1-cycle latency to help achieve a better performance. Filled pipelines are not shown in Figure \ref{fig:fig26}. If the accumulator can get the result in one cycle, then the speed of the three stages are the same and the pipeline is filled, which means the resource utilization is the best.

\subsubsection{Max-pooling Units}
Max-pooling consists of 8 parallel floating-point comparators and 1 FIFO. Its formula is as follows, in which $D$ stands for the elements of input matrix and $A$ stands for the elements of output matrix.

\begin{equation}
    A _{wo,co,ho} = \max_{\substack{(wo-1)\cdot s \leq w < (wo-1)\cdot s + k \\ (ho-1)\cdot s \leq h < (ho-1)\cdot s + k}} D _{w,h,co}
\end{equation}

Max-pooling does not change the input matrix channel. It only changes the surface size of input matrix, so the flow of computation is one-dimension less than convolution. Only Kernel, W, H, C are involved. (W and H can be exchanged.)

\begin{figure}
  \centering
  \includegraphics[width=1.0\textwidth]{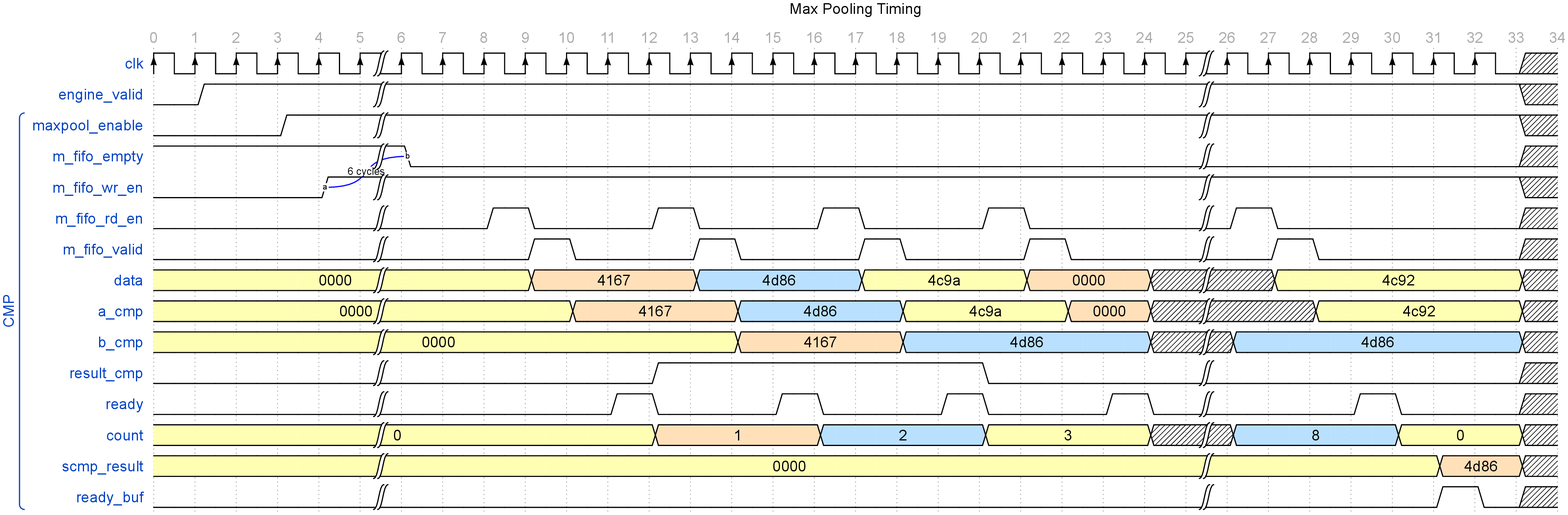} 
  \caption{Max-pooling timing sequence.}
  \label{fig:fig27}
\end{figure}

Figure \ref{fig:fig27} shows the microscopical RTL comparing process. The parallelism is 8. It is a one-stage flow. After \texttt{engine\_valid} is high, \texttt{maxpool\_enable} is pulled high and computation units are enabled. The computation units read out data from BRAM (not shown in Figure \ref{fig:fig27}). After 6 cycles \texttt{m\_fifo\_empty} is pulled low. 8 parallel comparators (initial value 0x0000) compare between new data from \texttt{M\_FIFO} and data in the previous cycle. If the result is high, which means new data in \texttt{a\_cmp} is larger than data in \texttt{b\_cmp}, value in \texttt{a\_cmp} will be replaced by that in \texttt{b\_cmp}. Otherwise \texttt{b\_cmp} remains unchanged. So \texttt{b\_cmp} stores the maximum value until the counter equals to \texttt{kernel}. The final result in \texttt{b\_cmp} is stored in \texttt{scmp\_result}, then \texttt{ready} signal is pulled high.

\subsubsection{Average-pooling Units}
Average-pooling consists of 8 parallel floating-point adders and 8 parallel floating-point dividers. Its computation formula is as follows, in which D stands for elements in the input matrix and A stands for elements in the output matrix.

\begin{equation}
    A _{wo,co,ho} = \frac{1}{k^2}\sum_{h=ho\cdot s}^{h=ho\cdot s + k} \sum _{w=wo\cdot s}^{w=wo\cdot s + k} D _{w,h,co}
\end{equation}

Average-pooling does not change the input matrix channel. It only changes the surface size of input matrix, so the flow of computation is one-dimension less than convolution, which is the same as max-pooling.

\begin{figure}
  \centering
  \includegraphics[width=1.0\textwidth]{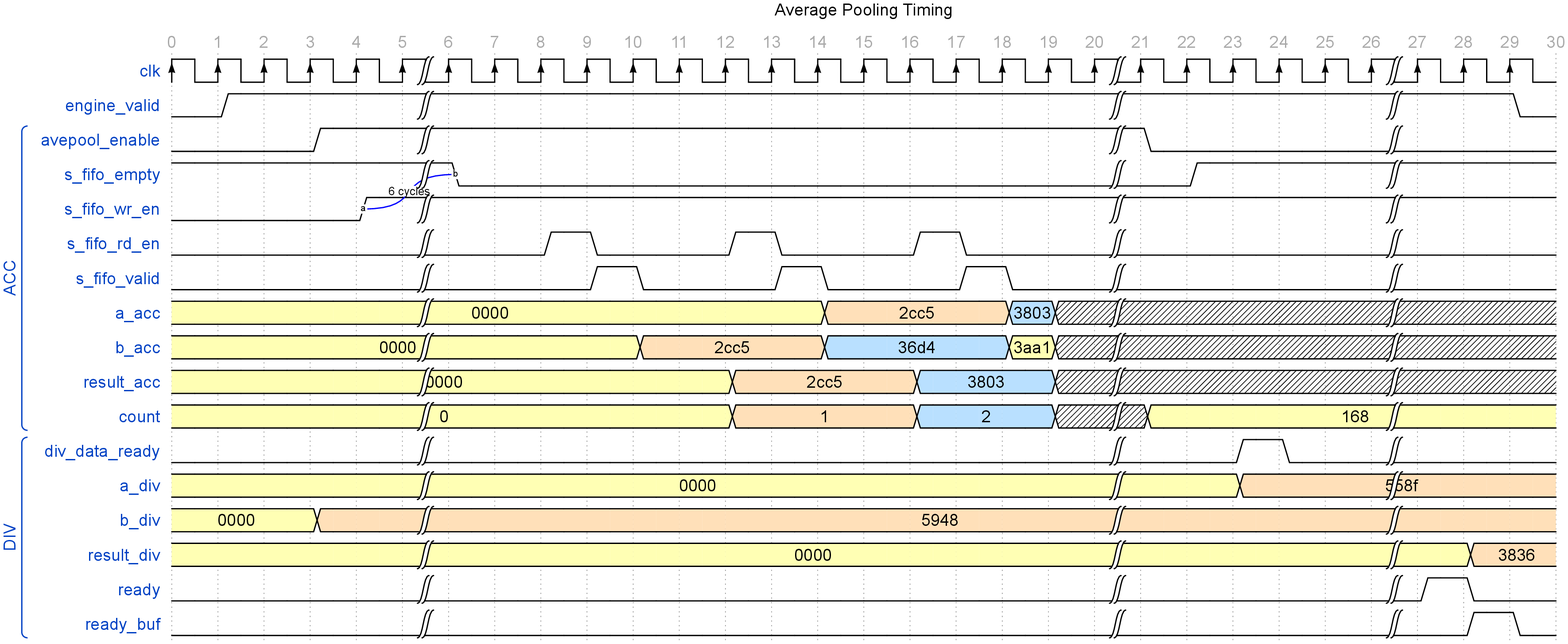} 
  \caption{Average-pooling timing sequence.}
  \label{fig:fig28}
\end{figure}

Figure \ref{fig:fig28} shows the microscopical RTL average-pooling process. The parallelism is 8. It is a two-stage flow. After \texttt{engine\_valid} is high, \texttt{average\_enable} is pulled high and computation units are enabled. The computation units read out data from BRAM (not shown in Figure \ref{fig:fig28}). After 6 cycles \texttt{s\_fifo\_empty} is pulled low. 8 parallel accumulators read data from \texttt{S\_FIFO} and do accumulation until the counter equals to \texttt{kernel}. Then \texttt{div\_data\_ready} will generate a one-cycle-wide pulse.

When \texttt{div\_data\_ready} is high, 8 parallel dividers will be triggered. \texttt{a\_div} stores the accumulating results from the previous stage. \texttt{b\_div} stores the int-FP converter output, which is \texttt{kernel} in FP16 format (In Figure \ref{fig:fig28}, it is 0x5948, i.e., $13\times 13 = 169$). After 6 cycles \texttt{ready\_buf} is pulled high and computation finishes.

\subsubsection{Test Files and Scripts}
\begin{figure}
  \centering
  \includegraphics[width=0.8\textwidth]{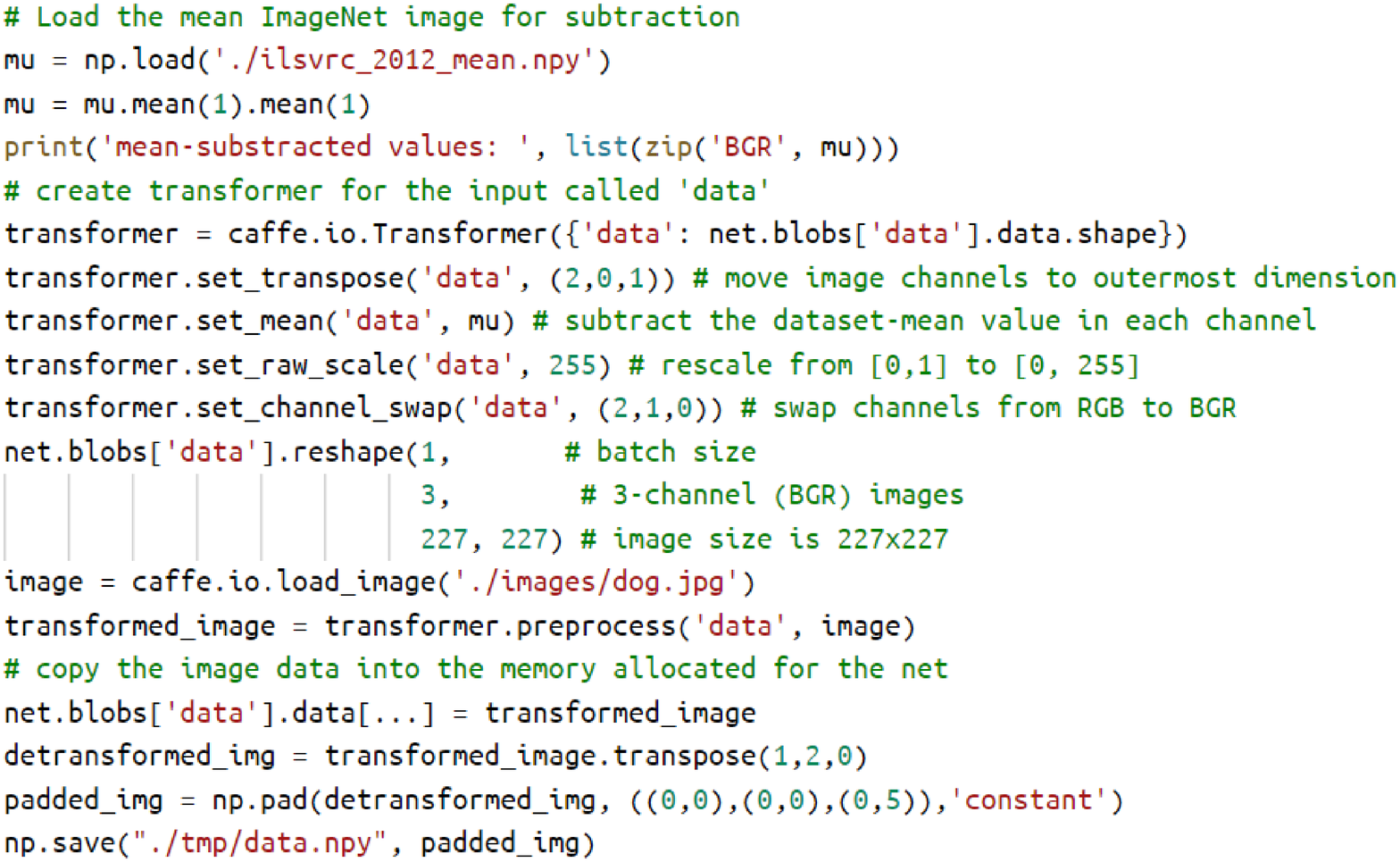} 
  \caption{Preprocess.py code script.}
  \label{fig:fig29}
\end{figure}

\begin{figure}
  \centering
  \includegraphics[width=0.45\textwidth]{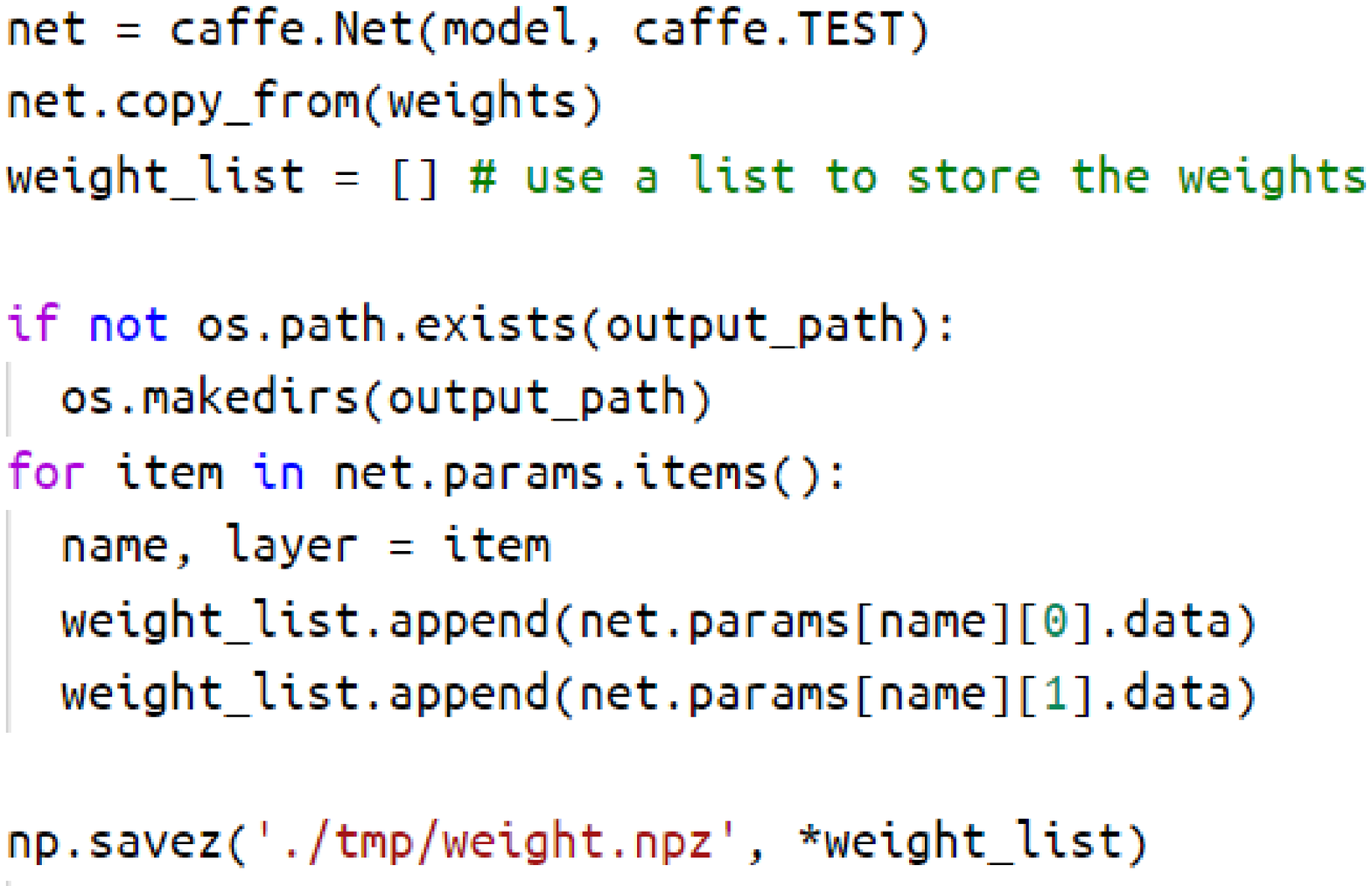} 
  \caption{Extract.py code script.}
  \label{fig:fig30}
\end{figure}

To generate the image data and weights required by CNN forwarding, there are some scripts to extract values in FP16 format from images and pre-trained models. Preprocess.py moves image channels to the lowest dimension, swap channels from RGB to BGR(Caffe), subtracts the mean value of ILSVRC\_2012 dataset in each channel from the image and re-scales the difference from [0, 1] to [0, 255] (Figure \ref{fig:fig29}). Extract.py extracts weights and biases from prototxt and caffemodel, converts them to FP16 format and pack them into npz format, as shown in Figure \ref{fig:fig30}. The host script calls the npz file in execution.

\begin{figure}
  \centering
  \begin{minipage}[c]{0.78\textwidth}
  \centering
  \includegraphics[width=1.0\textwidth]{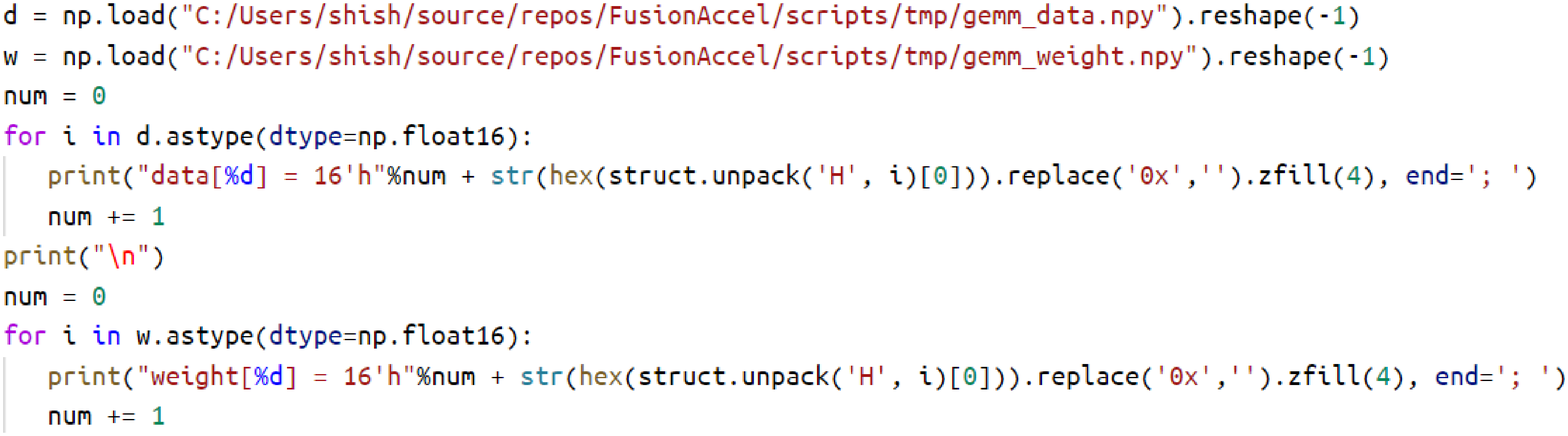} 
  \end{minipage}
  \vline
  \begin{minipage}[c]{0.2\textwidth}
  \centering
  \includegraphics[width=1.0\textwidth]{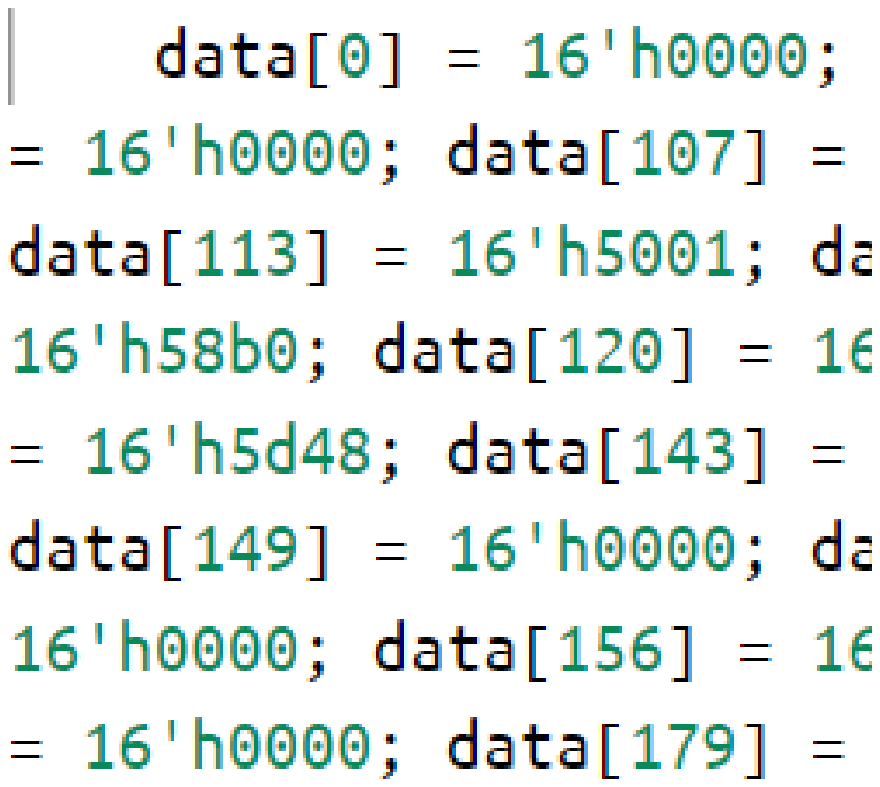} 
  \end{minipage}
  \caption{The testbench generated by python script.}
  \label{fig:fig31}
\end{figure}

Macros are used in computation testbenches to test if convolution, average-pooling, and max-pooling are correct. On the other hand, because the data volume is huge in the testbench, python scripts are used to generate testbench data from the pre-processed image and network weights, as shown in Figure \ref{fig:fig31}. Such script generates the testbench codes directly from Numpy matrix and saves manual input.

Apart from these scripts, Caffe on CPU is also required to verify the inference, which is identical to the BVLC sample script \footnote{\href{https://nbviewer.jupyter.org/github/BVLC/caffe/blob/master/examples/00classification.ipynb}{BVLC Classification: Instant Recognition with Caffe}. Accessed March 6, 2019.}.

\subsection{USB3.0 IO}
USB3.0 IO block loads input commands to command FIFO and stores input data, weight and bias to the corresponding cache. Meanwhile it transfers the result to result FIFO, and the parameters to host to calculate the cache positions.

\begin{figure}
  \centering
  \includegraphics[width=0.8\textwidth]{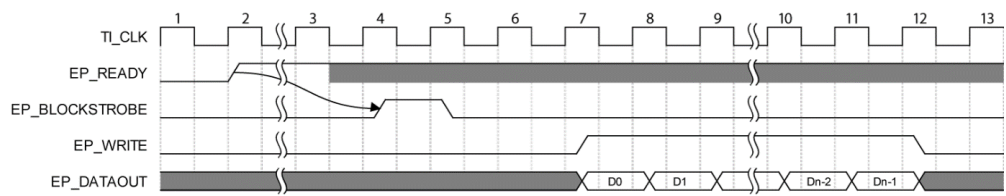} 
  \caption{Block-Throttled PIPE IN timing diagram.}
  \label{fig:fig32}
\end{figure}

\begin{figure}
  \centering
  \includegraphics[width=0.8\textwidth]{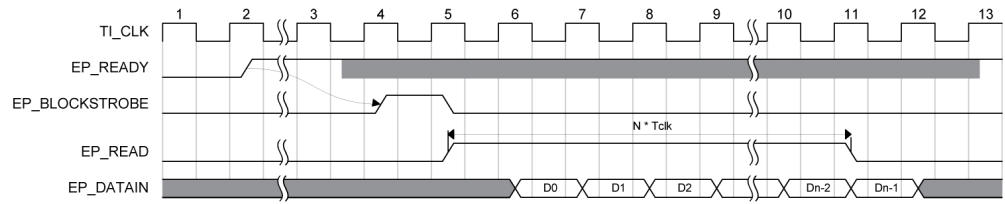} 
  \caption{Block-Throttled PIPE OUT timing diagram.}
  \label{fig:fig33}
\end{figure}

Utilized USB3.0 FrontPanel APIs \footnote{\href{https://docs.opalkelly.com/display/FPSDK/FrontPanel+SDK}{Opal Kelly FrontPanel SDK}. Accessed March 6, 2019.} are as follows. Wire In writes single 32-bit value to FPGA registers. Wire Out reads single 32-bit value from FPGA registers. Block-Throttled Pipe In is block write with handshake as timing Figure \ref{fig:fig32}. Block-Throttled Pipe Out is block read with handshake as timing Figure \ref{fig:fig33}. \texttt{EP\_READY} is enable signal of USB communications from FPGA, which is related to the available space of CMDFIFO and RESFIFO. \texttt{EP\_WRITE}/\texttt{EP\_READ} is the write and read signal sent directly by PC host API. \texttt{EP\_DATAOUT} and \texttt{EP\_DATAIN} are 32 bits wide.

\subsection{FIFOs and BRAMs}
Figure \ref{fig:fig22} shows that in the top-level hardware design, there are CMDFIFO on the input side and RESFIFO on the output side.

\begin{figure}
  \centering
  \begin{minipage}[t]{0.45\textwidth}
  \centering
  \includegraphics[width=1.0\textwidth]{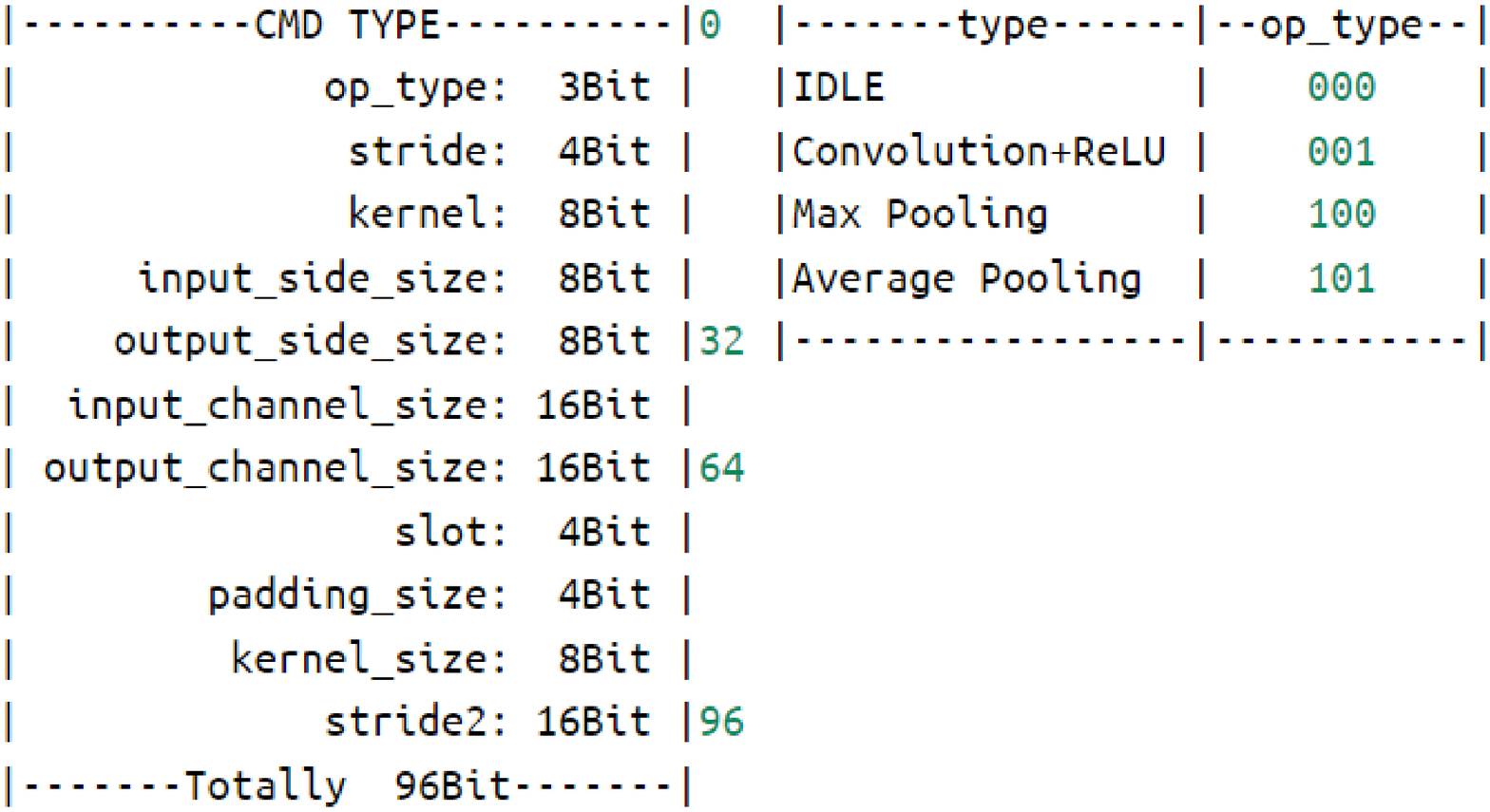} 
  \caption{Network parameters required by the accelerator.}
  \label{fig:fig34}
  \end{minipage}
  \begin{minipage}[t]{0.53\textwidth}
  \centering
  \includegraphics[width=1.0\textwidth]{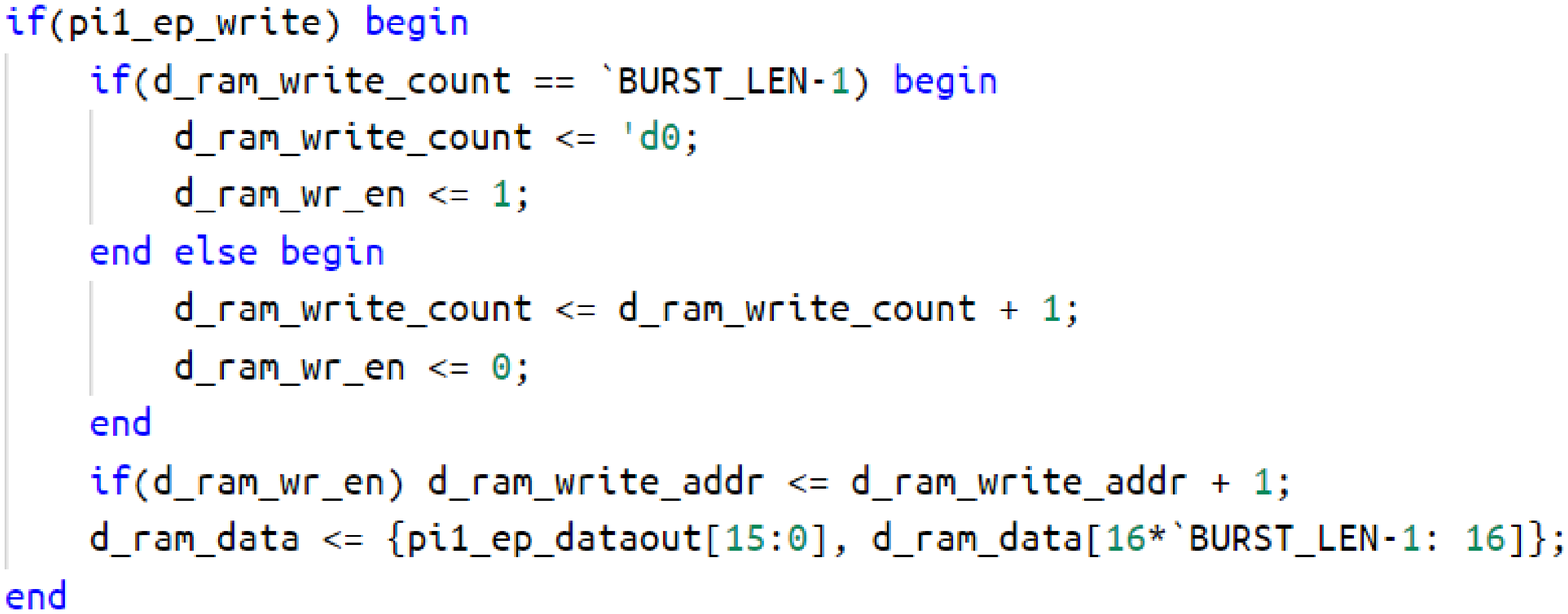} 
  \caption{SERDES before BRAM read.}
  \label{fig:fig35}
  \end{minipage}
\end{figure}

Command FIFO is 32 bits wide and 1024 deep. It stores parameters of each layer. Since each layer requires 12Bytes to characterize, as shown in Figure \ref{fig:fig34}, theoretically 341 layers are supported. If the network is deeper, we can just increase command FIFO depth.

\texttt{op\_type} is the computation format of this layer, idle, convolution + ReLU, max-pooling or average-pooling. \texttt{stride} ranges from 0 to 3. Large strides lose information of the image while small ones make the forwarding structure large, so 4 bits are enough. \texttt{kernel} is the side size of computation window. \texttt{kernel\_size} is the square size of computation windows, i.e., \texttt{kernel\_size = kernel * kernel}. This extra parameter saves integer multiplication on-chip, while the resource and communication time is trivial. \texttt{input\_side\_size} is the side of input matrix ($w = h$), for example the parameter of the first layer in SqueezeNet v1.1 is 227. \texttt{output\_side\_size} is the side of output surface($w = h$), for example the parameter of the first layer in SqueezeNet v1.1 is 113.

\texttt{input\_channel\_size} is the channel of input matrix. \texttt{output\_channel\_size} is the channel of output matrix. In convolution layer normally \texttt{input\_channel\_size} is smaller than \texttt{output\_channel\_size}. This is because with the convolution is a process that the network gets deeper, and surface becomes smaller. For pooling layer, \texttt{input\_channel\_size} equals to \texttt{output\_channel\_size}.

\texttt{padding\_size} is normally 1, but 4 bits are still reserved for it. \texttt{stride2 = stride * kernel}. Such value will be called repeatedly in computation. As a network parameter it reduces logic utilization and helps timing convergence.

\texttt{slot} determines whether this layer belongs to any of parallel layers, like expand1x1 and expand3x3 in SqueezeNet v1.1. The input cube of these two layers comes from the output of same layer, and the outputs of the two layers are merged into one matrix as input to the next layer. \texttt{slot[0]} and \texttt{slot[1]} indicates the order of two parallel layers, and \texttt{slot[3:2]} shows the total number of parallel layers. \texttt{slot} is only transferred to PC host to help parse the input matrix and not called by computation units, because in practical computation these two parallel layers are in sequential orders.

Result FIFO is 32 bits wide, and 1024 deep. It directly stores the computation result and transfers it via USB3.0 upon host request. We can easily get that under channel-first parallelism such FIFO can support result convolution layer of a maximum \texttt{input\_side\_size} of 1024 (since a single layer corresponds to output channel of 1) or pooling layer whose \texttt{input\_side\_size} is 128 (since output channel parallelism is 8). This is enough for concurrent networks. If the side or area of intermediate layers is larger, result FIFO depth should be enlarged.

In top-level hardware design, as shown Figure \ref{fig:fig22}, there are three BRAM as caches, data cache, weight cache and bias cache. These BRAM caches will be accessed once in every cycle to extract value to the corresponding registers of the same width.

Data cache is 128 bits wide, and 1024 deep. It serves to store the convolution, average-pooling and max-pooling data received via USB3.0. Since these three computation modules are operating only when \texttt{cmac\_enable}, \texttt{avepool\_enable}, and \texttt{maxpool\_enable} are high, so it is not necessary to add MUX on the data path. Weight cache is 128 bits wide, and 8192 deep. It will be accessed by convolution only. Since the input channel parallelism is 8 in convolution, and the data format is FP16, so the width is 128 bits. If the max kernel size in convolution is $3\times 3 = 9$, then the max supported \texttt{input\_channel\_size * output\_channel\_size} is $c\times n = 8192\times 8\div 9 = 7281$. Since the output channel parallelism is also 8, the max \texttt{input channel size} is $c = 8192\div 9 = 910$, which meets the requirement of SqueezeNet v1.1 (as in Table \ref{tab:table2}).

Bias cache is 128 bits wide, and 1024 deep. It will be accessed by convolution only. In FP16 format, only the lowest 16 bits are valid data in each bias cache, and the rest bits are all zeros. Since the output channel parallelism is 8 at most in convolution computation in this project, a bias cache depth of 8 should be enough. But to simplify the project files, same BRAM as data cache is utilized.

As for data cache and weight cache, since the USB3.0 input is 32 bits wide, and only the lower 16 bits are valid in FP16 format, so in every $128\div 16 = 8$ cycles a group of parallel data is cached by SERDES, as shown in Figure \ref{fig:fig35}. When the count is smaller than \texttt{BURST\_LEN - 1 = 7}, 16-bit data is shifted in until the 128-bit cache is filled.

\begin{figure}
  \centering
  \includegraphics[width=0.7\textwidth]{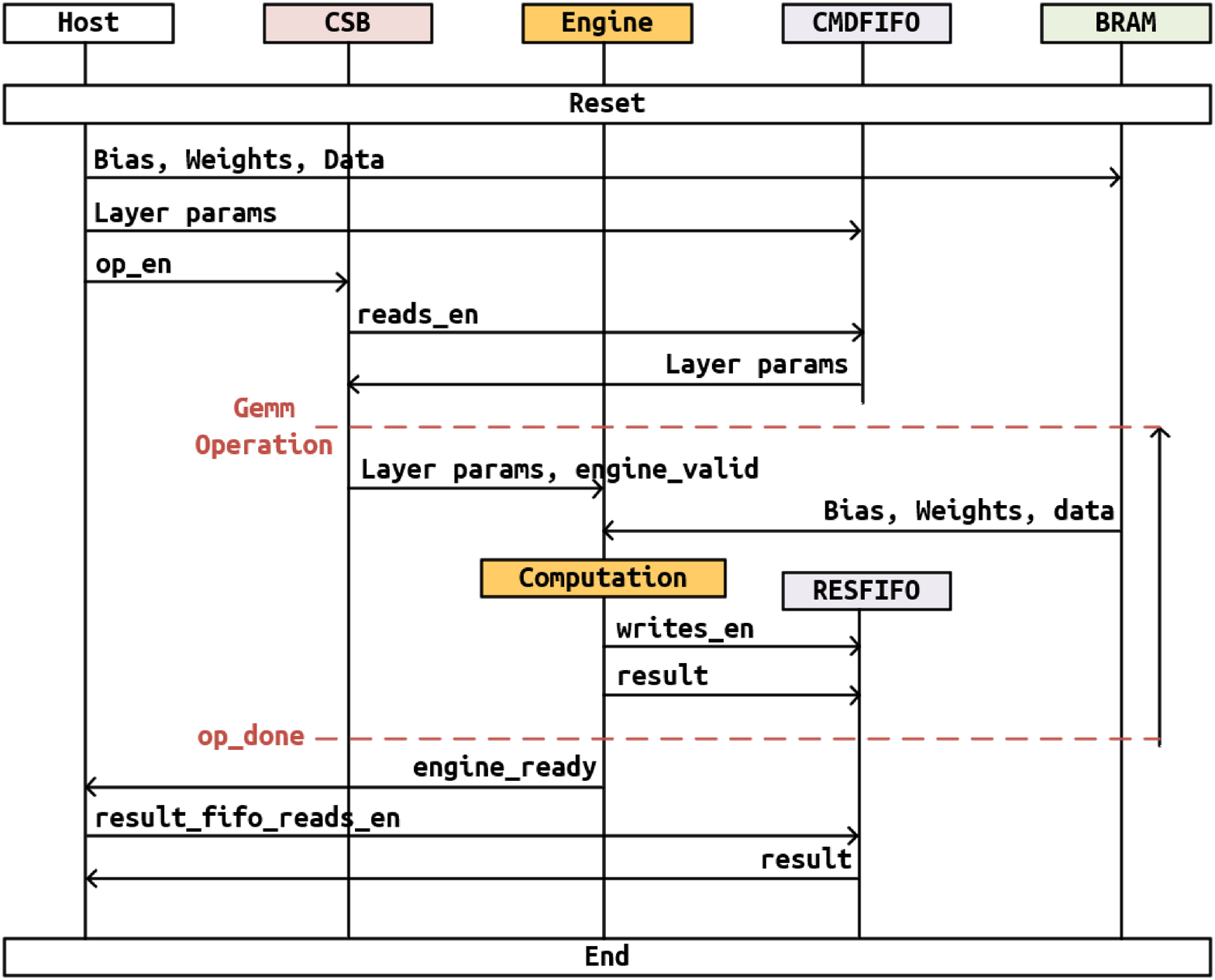} 
  \caption{Operation flow of the stream accelerator.}
  \label{fig:fig36}
\end{figure}

In Figure \ref{fig:fig36}, after resetting, all parameters will be initially loaded to CMDFIFO, while bias, weight and data will be transferred to BRAM by layer and by piece. After receiving \texttt{engine\_valid} high, the computation unit will read bias, data and weight from BRAM to calculate the result, and then write to RESFIFO. After this piece is finished, the host will get the result from RESFIFO via USB3.0 and the interrupt signal. Then goes the next piece, and next layer till the entire network forwarding is finished.

Compared to generic accelerator workflow as Figure \ref{fig:fig15}, stream accelerator workflow is relatively simplified, the data stack is thinner and the data path is shorter, which helps improve the performance and scale the parallelism.

\section{Software, System Architecture and Results}
\begin{figure}
  \centering
  \includegraphics[width=0.8\textwidth]{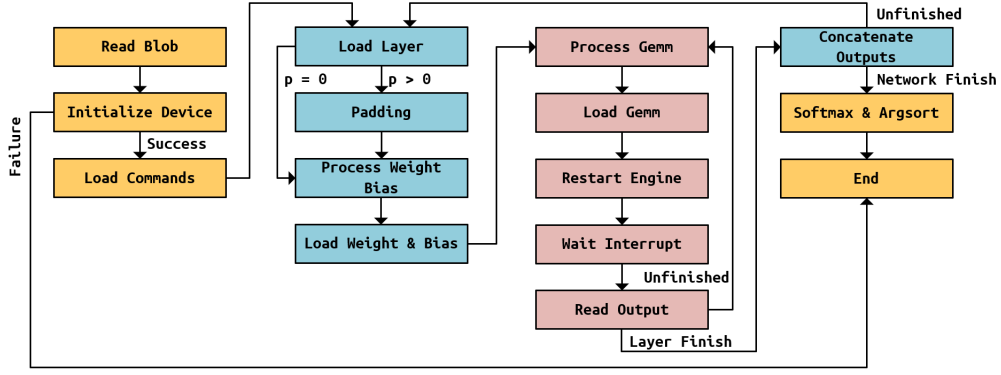} 
  \caption{Flow diagram of the software.}
  \label{fig:fig37}
\end{figure}

Figure \ref{fig:fig37} shows the PC host workflow. Detailed execution process in every stage is as follows. In \emph{Read Blob}, the host will load the network parameters, biases and weights from the generated packed npz by extract.py, and load the pre-processed image data by preprocess.py. In \emph{Initialize Device} the host connects and initializes FPGA and downloads the bitstream file. In \emph{Load Commands} the host transfers all parameters of each layer to CMDFIFO on FPGA. In \emph{Load Layer} these pre-stored parameters will be read out. These parameters will be called by computation units, as well as be used to slice the data blocks. In \emph{Process Weight Bias} the network weights will be processed and slices. In load weight \& bias the biases and weights will be transferred to bias cache and weight cache on FPGA.

In \emph{Process Gemm} the host slices the padded data block. These sliced data will be transferred to FPGA in \emph{Load Gemm}. \emph{Restart Engine} resets the computation unit, clear the registers after the calculation of the previous layer, and starts the computation unit. After all results are calculated, FPGA will notify the host via interrupt. Then the host will fire a read request. In \emph{Read Output} the data will be transferred from RESFIFO. In \emph{Concatenate Outputs}, all pieces of results will be concatenated to be the input of the next layer. Finally \emph{Softmax \& Argsort} will normalize and sort the final result. Softmax function formula is as follows:

\begin{equation}
    \sigma (D_i) = \frac{e^{D_i}}{\sum _{k=0}^{k=K-1}e^{D_k}}, for\ i = 0,1,2,...,K-1
\end{equation}

It serves to normalize the output value to (0, 1), this value stands for the possibility the input image is the ith item inferred by the network.

Table \ref{tab:table3} is the resource utilization of the accelerator after synthesize, placement and routing. We can see that there are still redundant registers, LUTs and DSPs. The reason why DSPs are not used much is that in Xilinx Floating Point 5.0 IP\footnote{\href{https://www.xilinx.com/support/documentation/ip_documentation/floating_point_ds335.pdf}{Xilinx DS535 Floating-Point Operator v5.0 LogiCORE Product Guide}. Accessed March 6, 2019.}, only multipliers use DSPs and the rest use Flip Flops and LUT resources. Also, we can see that RAMB16BWERs are almost used up. This is because in the design many BRAMs and FIFOs are used. LUT usages are mostly made up of floating-point computation units. LUT utilization are over 70\% when the parallelism is 16. A doubled parallelism means doubled width in BRAM and FIFO because of channel-first parallelism. However, the present RAM16BWER and RAMB8BWER utilization exceeds 50\%, so this chip is not capable of holding parallelism of 16.

\begin{figure}
  \centering
  \includegraphics[width=0.8\textwidth]{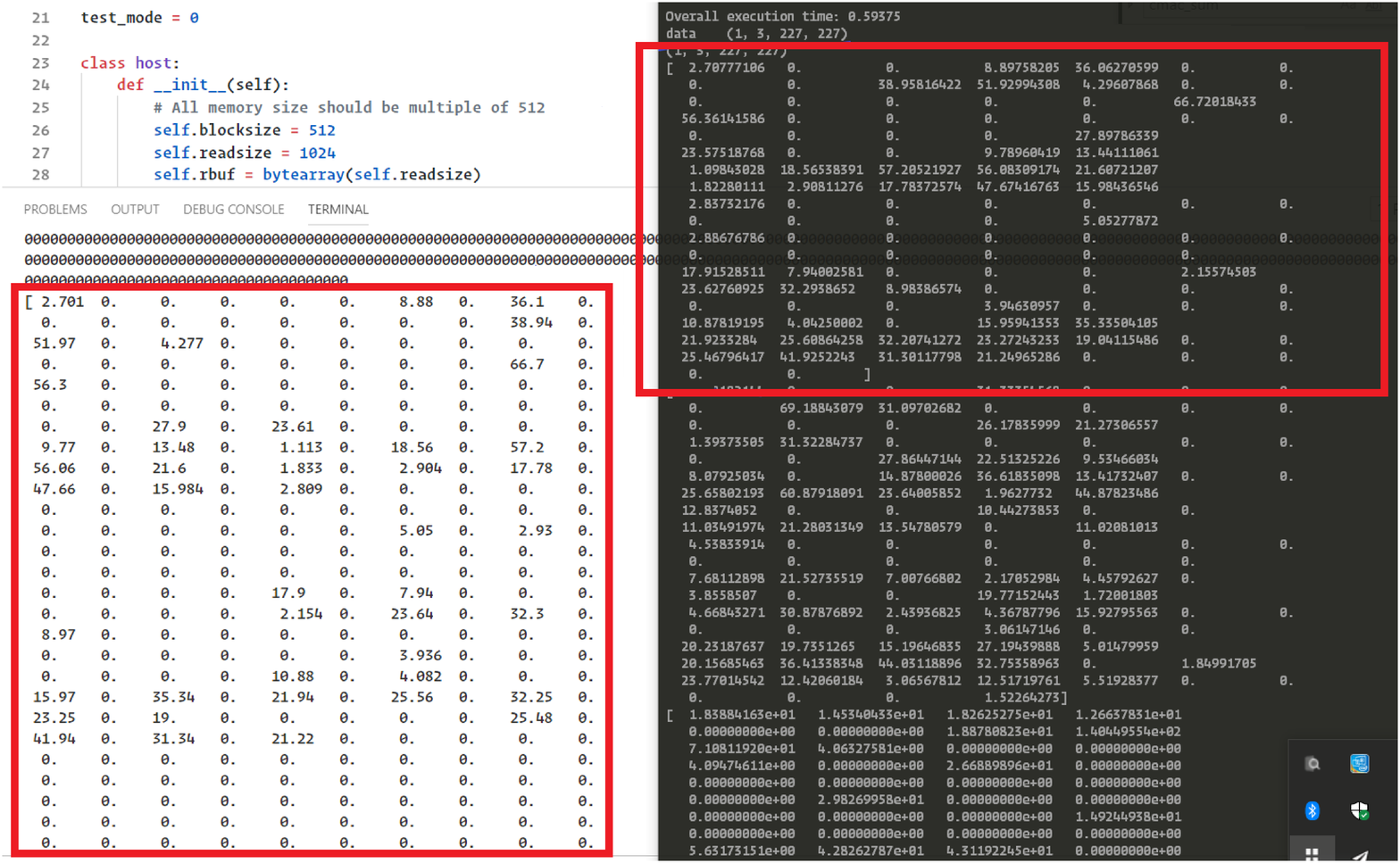} 
  \caption{Intermediate result of the accelerator computation.}
  \label{fig:fig38}
\end{figure}

Figure \ref{fig:fig38} is the forwarding result (left) of the first layer of SqueezeNet v1.1 compared to that (right) on Caffe CPU. Since the accelerator is in FP16 format, between every two results there is a padded 0. We can see that the results on these two platforms are basically the same, and deviations just start from the second or third decimal place.

\begin{table}
  \centering
  \caption{Logic resource utilization of the FPGA.}
  \includegraphics[width=0.7\textwidth]{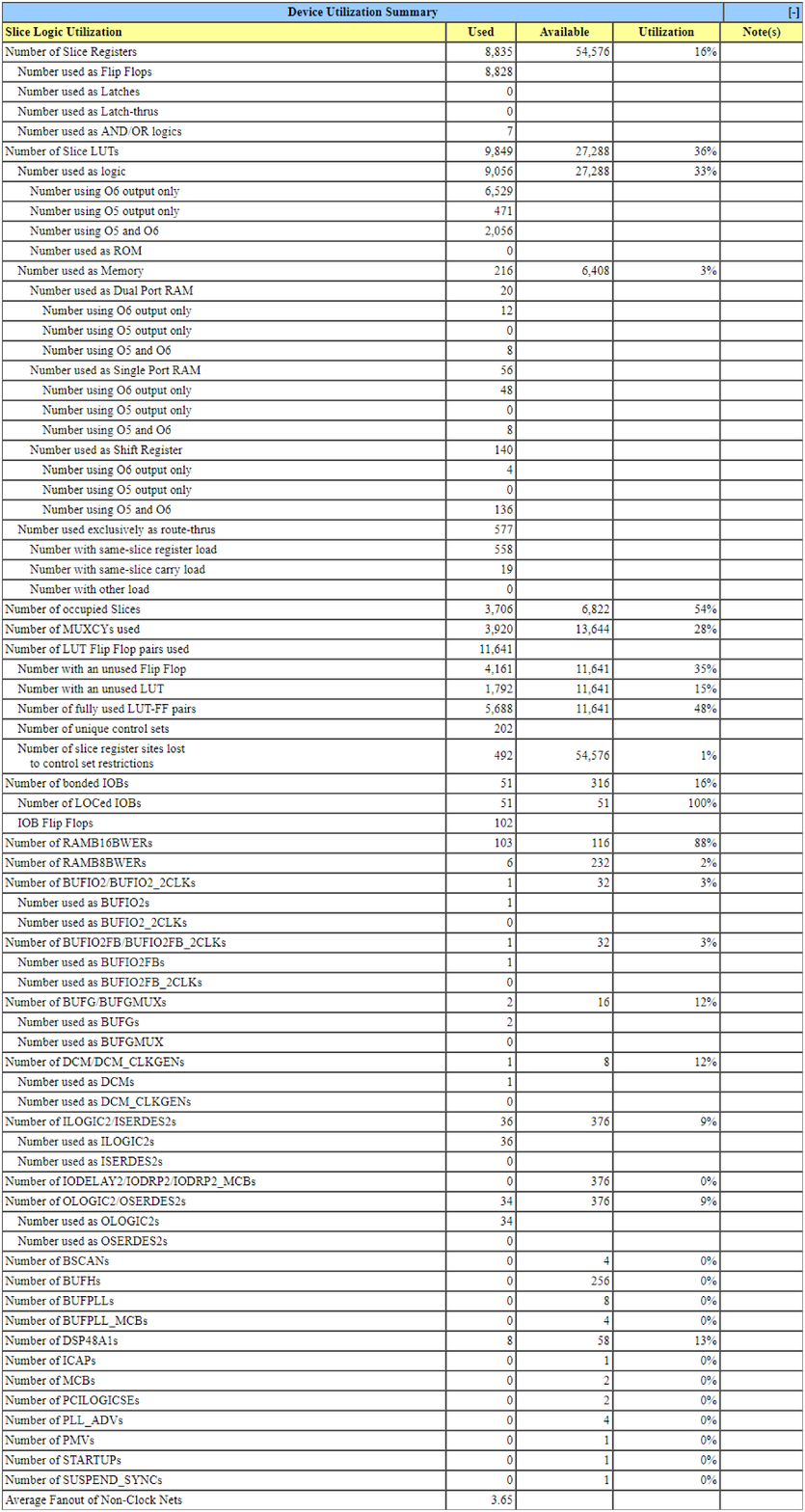} 
  \label{tab:table3}
\end{table}

\begin{figure}
  \centering
  \includegraphics[width=0.8\textwidth]{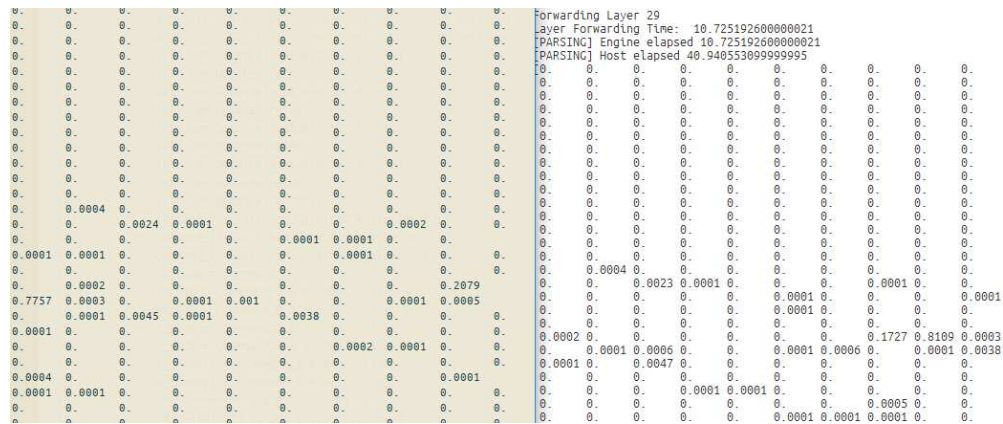} 
  \caption{Final result of the accelerator computation.}
  \label{fig:fig39}
\end{figure}

\begin{figure}
  \centering
  \begin{minipage}[t]{0.28\textwidth}
  \centering
  \includegraphics[width=1.0\textwidth]{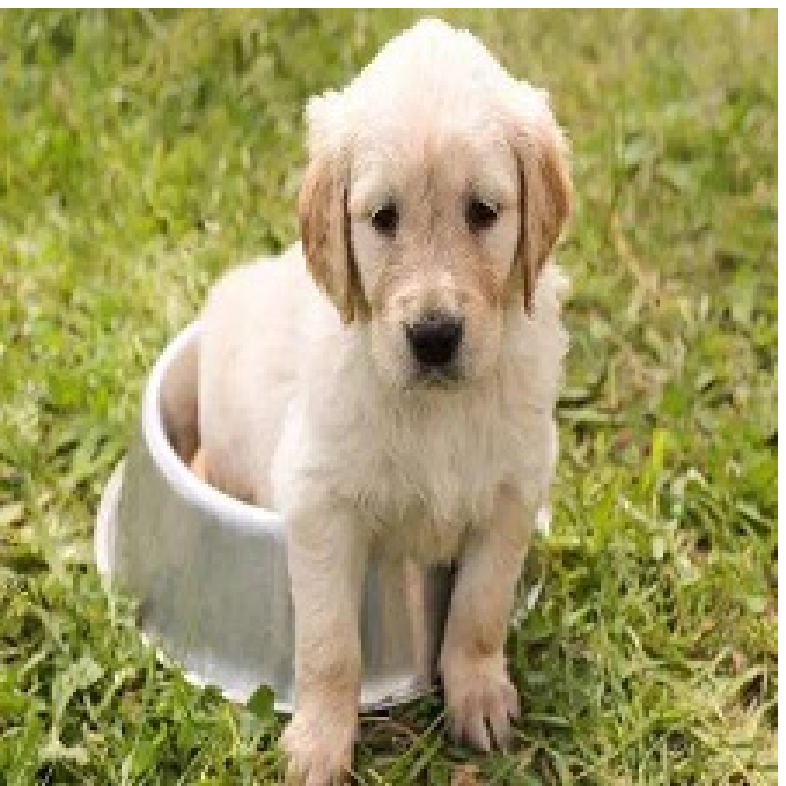}
  \end{minipage}
  \begin{minipage}[t]{0.7\textwidth}
  \centering
  \includegraphics[width=0.9\textwidth]{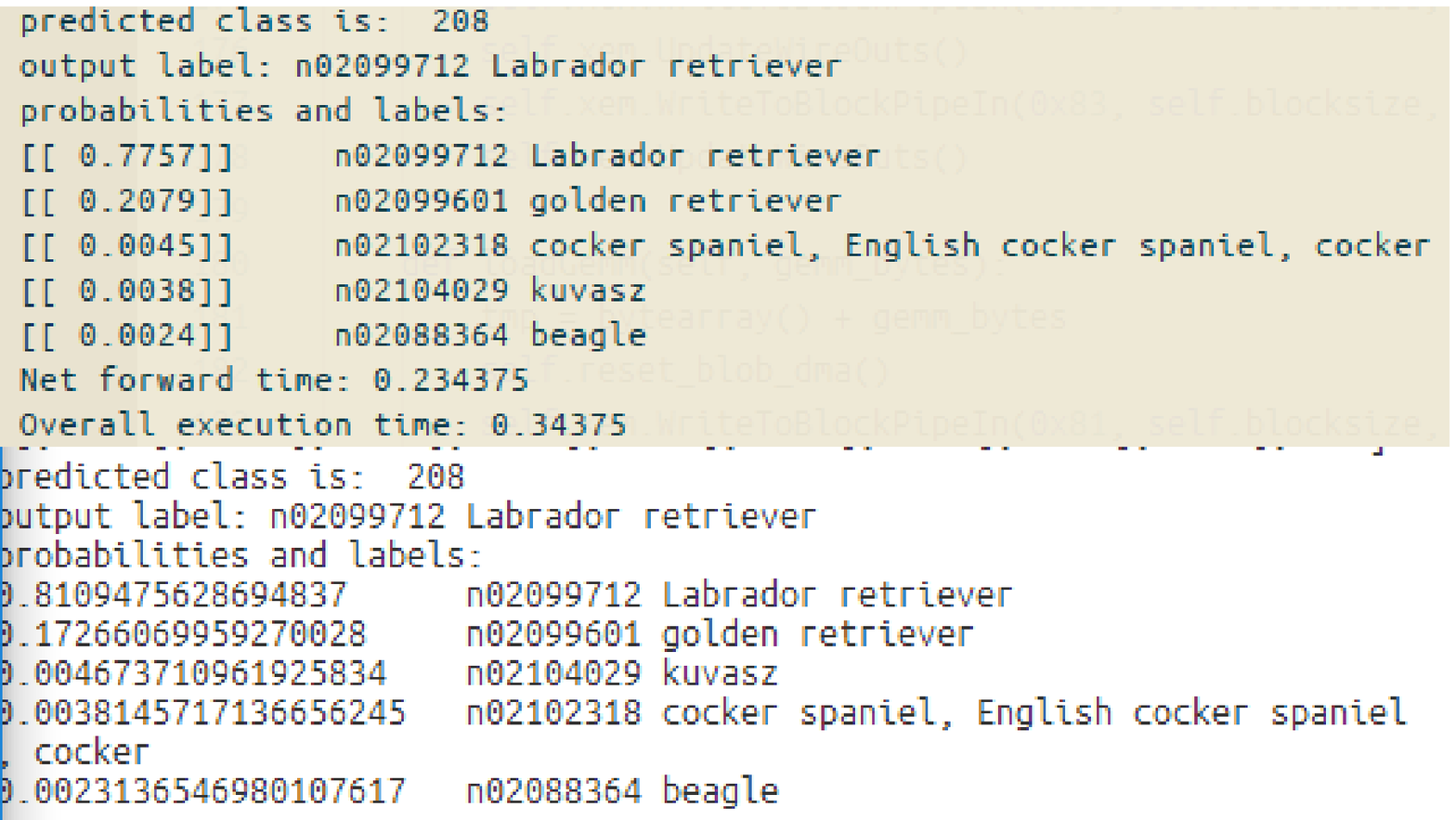} 
  \end{minipage}
  \caption{Caffe inference result (upper) and accelerator inference result.}
  \label{fig:fig40}
\end{figure}

Figure \ref{fig:fig39} and \ref{fig:fig40} shows that FPGA results are identical to Caffe CPU results. FP16 does not bring about any errors or obvious deviations because softmax amplifies the result of the final-layer convolution. Larger numbers on the exponential are more significant, but this does not compromise the correctness of the result much. Because the hardware resource restrictions (RAM16BWER utilization is already 88\%), the parallelism is only 8. Hence, the computation elapse time is relatively long (computation time is 10.7s, and the whole process is 40.9s), which is significantly slower than Caffe. If USB3.0 can be replaced by PCIe buses, the latency will be improved. If there are more hardware resource to improve parallelism, the computation time will be proportionally reduced.

\section{Conclusion and Future Development}
\subsection{Conclusion}
This paper analyzes the optional solutions and architectures of CNN accelerators, as well as some trade-offs between optional algorithms to preserve the scalability of the hardware project. A stream processing architecture with im2col + GEMM convolution solution is eventually designed. Channel-first parallelism is used to simplify the computation control logic. Re-configurable accelerator hardware and corresponding host software are generated.

In the verification stage, SqueezeNet v1.1 is used to test the inference sanity on FPGA, and the result is identical to that on Caffe CPU. The only deviation is introduced by the precision difference between FP16 and FP32. Since the scalability is considered in design, the selected algorithms maximize parallel computing resources, and the computation resource overhead is not influenced by network types or number of layers.

Restricted by FPGA resources, the speed of the accelerator is slower than CPU and other concurrent accelerators. If FPGAs with more logic resources are used, if parallelism are improved when this project is migrated to ASIC, or if higher clock speeds and lower-latency high-speed buses are used, better accelerator performance can be achieved.

\subsection{Configurable Parameters towards ASIC \& Optimization}
\begin{figure}
  \centering
  \includegraphics[width=0.25\textwidth]{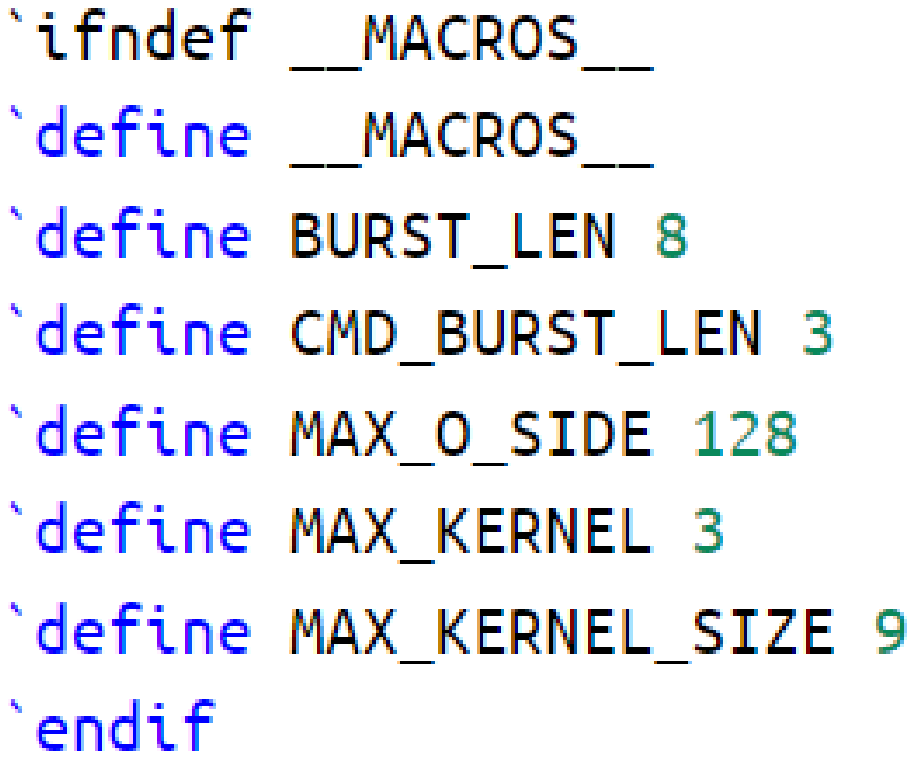} 
  \caption{Configurable parameters before compilation.}
  \label{fig:fig41}
\end{figure}

In this project, the computation precision and parallelism are two most important configurable parameters. These two parameters determine the numbers of computation units and the width of caches and FIFOs. In the current design the precision is FP16 and parallelism is 8. These parameters can be adjusted by macros, as shown in Figure \ref{fig:fig41}. The reason why such scaling is practical is that channel-first parallelism is used in design and no other logic must be changed while scaling. Apart from parameters above, \texttt{MAX\_KERNEL} and \texttt{MAX\_O\_SIDE} are also configurable. These two determines the overhead of RAMs as buffer of results. \texttt{CMD\_BURST\_LEN} determined how many double words (4Bytes) are read out from CMSFIFO to CSB. In this case, the number is 3, i.e., 12Bytes.

If the project is going to be migrated to ASIC, we need to replace some FPGA IPs including floating-point IP, USB3.0 IP, BRAM IP and FIFO IP. There are practical ASIC solutions for these IPs, and thanks to handshake protocols, the core logic does not have to change much, just matching the ASIC IP.

Since the hardware in the project uses an engine to compute the CNN forwarding rather than storing weights directly on hardware, and the scale of computation units are not related to the intrinsic parameters of networks, other networks like AlexNet are also supported. Thus, this project is configurable in runtime. On larger FPGAs or on ASICs, more computation units (e.g., in NVDLA full configuration there are totally 2048 parallel convolution multiply-accumulators) and SRAMs can be used to boost up the forwarding process. On the other hand, the host logic can also be migrated to CPUs like ARM of RISC-V.

Moreover, the network parameters are manually extracted rather than by script. This is because the parameter requirements change during the design process. After the architecture is fixed, the commands can by extracted from prototxt by python script.

\nocite{*}
\bibliographystyle{unsrt}  

\end{document}